\documentclass[%
reprint,
aps,
pra
]{revtex4-2}

\usepackage[usenames]{color}
\usepackage{colortbl}
\usepackage{graphicx}
\usepackage{dcolumn}
\usepackage{bm}
\usepackage{amsmath}
\usepackage[utf8]{inputenc}

\usepackage{comment}

\begin{document}

\preprint{APS/123-QED}

\title{Energy Transport Induced by Transition from Weak to Strong Coupling Regime Between Non-Hermitian systems}

\author{I. V. Vovcenko}
\affiliation{%
 Moscow Institute of Physics and Technology, 9 Institutskiy pereulok, Dolgoprudny 141700, Moscow region, Russia;
}%
\affiliation{%
Kotelnikov Institute of Radioengineering and Electronics, Mokhovaya 11-7, Moscow, 125009, Russia
}%


\author{A. A. Zyablovsky}
\affiliation{
 Dukhov Research Institute of Automatics (VNIIA), 22 Sushchevskaya, Moscow 127055, Russia;
}
\affiliation{
 Moscow Institute of Physics and Technology, 9 Institutskiy pereulok, Dolgoprudny 141700, Moscow region, Russia;
}%
\affiliation{
 Institute for Theoretical and Applied Electromagnetics, 13 Izhorskaya, Moscow 125412, Russia;
}
\affiliation{%
Kotelnikov Institute of Radioengineering and Electronics, Mokhovaya 11-7, Moscow, 125009, Russia
}%

\author{A. A. Pukhov}
\affiliation{
 Moscow Institute of Physics and Technology, 9 Institutskiy pereulok, Dolgoprudny 141700, Moscow region, Russia;
}%
\affiliation{
 Institute for Theoretical and Applied Electromagnetics, 13 Izhorskaya, Moscow 125412, Russia;
}

\author{E. S. Andrianov}
\email{andrianov.es@mipt.ru}
\affiliation{
 Dukhov Research Institute of Automatics (VNIIA), 22 Sushchevskaya, Moscow 127055, Russia;
}
\affiliation{
 Moscow Institute of Physics and Technology, 9 Institutskiy pereulok, Dolgoprudny 141700, Moscow region, Russia;
}
\affiliation{
 Institute for Theoretical and Applied Electromagnetics, 13 Izhorskaya, Moscow 125412, Russia;
}
\date{\today}

\begin{abstract}
Recently, strong coupling between non-Hermitian physical systems of different nature is widely investigated due to it endows them with new properties. 
In this work, we investigate the energy transport between strongly coupled  systems.
We use a partial-secular approach for the description of an open quantum system to investigate the system dynamics during the transition from a weak to a strong coupling regime with an increase of coupling between subsystems.
On the example of strongly coupled two-level atoms, we show that near the transition point enhancement of energy transport between the system and reservoirs takes place.
This manifests in the fact that energy flow normalized to the coupling constant reaches the maximum both in the cases of zero and non-zero frequency detuning.
We show that maximization of normalized energy flow can be used for the determination of the transition to the strong coupling regime in the case of non-zero detuning when there is no clear transition point from the weak to strong coupling regime.
The suppression of the energy flow at high relaxation is demonstrated.

\end{abstract}

\maketitle

\section{Introduction}
In the last decades, strong coupling between non-Hermitian physical systems has been actively investigated due to its importance from both fundamental and practical points of view \cite{Miri_Alu, ozdemir2019parity, hummer2013weak, torma2014strong, yoshie2004vacuum, hennessy2007quantum, chikkaraddy2016single, zengin2015realizing, munkhbat2018suppression, reithmaier2004strong}.
A strong coupling regime is achieved when an interaction constant between subsystems is larger than relaxation rates \cite{chikkaraddy2016single}; in the opposite case, a weak coupling regime is realized.
The strong coupling allows for increasing entanglement time between qubits even in the case of large dephasing ~\cite{DAE}, to control the rate of the chemical reaction when molecules are strongly coupled to plasmonic resonators \cite{galego2016suppressing,munkhbat2018suppression,flick2018strong,nefedkin2020role,doronin2021resonant}, to achieve laser generation without inversion \cite{doronin2019lasing,doronin2021strong}.
The strong coupling regime is separated from the weak coupling regime by an exceptional point (EP) - the point in the space of the system parameters at which two or more system eigenstates become linearly dependent and their eigenfrequencies coincide \cite{Miri_Alu,ozdemir2019parity,moiseyev2011non,berry2004physics}.
Operation near the EP makes it possible to enhance sensitivity of sensors ~\cite{chen2017exceptional,hodaei2017enhanced} and to improve operation of laser gyroscopes ~\cite{lai2019observation}.

Interest to strong coupling regime is also related to control of energy transport in open quantum systems \cite{yang2020phonon, sergeev2021new}. 
In \cite{sergeev2021new}, it has been demonstrated that in strong coupling regime a new type of phase transition appears. 
This phase transition manifests itself as an appearance of two maxima in Fourier spectrum of energy flow from subsystem connected with hot reservoir to subsystem connected with cold reservoir \cite{sergeev2021new}. 
Changing of the temporal dynamics of energy flows also takes place in strongly coupled optomechanical systems~\cite{yang2020phonon}.

EPs exist in open non-Hermitian systems which interact with external reservoirs. 
One of the possible ways to describe their dynamics is to use the Born-Markov approximation which assumes that the reservoirs at all moments of time are in thermal equilibrium, and the system dynamics is local in time.
These approximations result in master equation in the Lindblad form for a system density matrix \cite{DaviesMME, LindbladMME, B-P}.
If the system consists of several coupled subsystems, e.g., two-level systems (TLSs) or oscillators, the relation between coupling constants and relaxation rates is crucial for the derivation of the master equation.
If coupling constants are much less than relaxation rates, it is convenient to use perturbation theory for derivation of the master equation \cite{LOC2,LOC3}. 
An elementary approach is to consider only zero order of the perturbation theory with respect to the coupling constants.
This approach is called local and supposes that the coupling does not affect the relaxation of an individual subsystem \cite{LOC1}.
On the other hand, if coupling constants are much greater than relaxation rates, the Born-Markov approximation together with the secular approximation can be used~\cite{DaviesMME, GLB1,GLB2}.
This approach is called global and assumes that relaxation depends on the eigenstates of the whole system.
It results in Gorini-Kossakovski-Sudarshan-Lindblad (GKSL), or Lindblad, master equation.
Near the EP coupling constants and relaxation rates are of the same order, so both local and global approaches are not applicable \cite{LOCvsGLB3,LOCvsGLB4}.
A suitable approach that is applicable near the EP is the partial-secular (PS) approach that has no restrictions on the relation of coupling constants and relaxation rates \cite{PS1,PS2,PS3}.

For the master equation in the Lindblad form one can justify thermodynamics laws \cite{kosloff2013quantum, Spohn1978, Spohn1978_convex}. 
Thermodynamics laws impose restrictions on the energy flow from the system to reservoirs.
In the stationary state, the first law of thermodynamics requires that the energy flow from the hot reservoir to the system equals the energy flow from the system to the cold reservoir.
Second law of thermodynamics described by Clausius inequality restricts entropy production and sign of energy flow.
Entropy production in the stationary state is zero, and second law of thermodynamics requires that the energy flow from the system to the cold reservoir should be a positive while from the system to the hot reservoir should be negative (i.e., energy flows from the hot reservoir to the system).
Both of these laws by themselves do not restrict absolute values of the energy flow that depend on the system state.
In \cite{LG}, the dependence of stationary energy flows on coupling constant between subsystems has been studied in the local and global approaches.
It has been shown that for large coupling constants stationary energy flows coincide in both approaches, but for small coupling constants local and global approaches predict very different stationary energy flows.
However, the influence of the EP on the energy flow has not been studied.
Moreover, the influence of coupling constant in PS approach and the impact of the detuning on system dynamics have not been investigated.

In this work, we study the dependence of the stationary energy flows between an open quantum system and reservoirs on the coupling constant between subsystems and coupling constants between system and reservoirs. 
We consider the cases of two coupled two-level subsystems.
Each of the subsystems is coupled with its reservoir with a given temperature.
The reservoir temperatures are supposed to be different.
We show that during the transition from the weak to the strong coupling regime the flow from the hot reservoir to the system and the flow from the system to the cold one tend to their maxima by absolute values.
Thus, the transition to the strong coupling regime leads to the enhancement of energy transport from the system to the reservoirs.
We demonstrate that specific energy flow, i.e., the energy flow divided by the coupling constant between subsystems, reaches a maximal value at some coupling constant between subsystems which is comparable with the coupling constant corresponding to the EP.
Hence, the maximization of specific energy flow can serve as an attribute of the transition from the weak to the strong coupling regime.
We show that in the case of non-zero detuning maximization of specific energy flow takes place at a non-zero coupling constant.
The last can serve as an attribute of transition to the strong coupling regime in the case of non-zero detuning when it is impossible to determine a transition to the strong coupling via the system spectrum.

\section{The model}
The Hamiltonian of an open quantum system interacting with reservoirs can be written as \cite{B-P}
\begin{equation}\label{TotHam}
\hat{H}=\hat{H}_S+\hat{H}_R+\hat{H}_{SR}.
\end{equation}
Here $\hat{H}_S$ is the Hamiltonian of the system, $\hat{H}_R$ is the Hamiltonian of the reservoirs and $\hat{H}_{SR}$ describes the interaction between the system and reservoirs.

In the Born-Markov approximation it is possible to eliminate reservoir degrees of freedom and obtain the master equation for the system density matrix in the Lindblad form ($\hbar=1$) \cite{DaviesMME, LindbladMME}:
\begin{gather}\label{LindGen}
	\frac{\partial \hat \rho}{\partial t} = -i\left[ \hat H_{\rm{S}}, \hat \rho \right] + \sum_j \hat L_j \left[\hat \rho \right].
\end{gather}
where $\hat L_j \left[\hat \rho\right]$ are Lindblad superoperators.
The concrete form of the Lindblad superoperators $\hat L_j\left[\rho\right]$ describing relaxation processes depends on the relation between relaxation rates and system eigenfrequencies.
This dependence is most pronounced in the case when the system consists of two interacted subsystems, and the system Hamiltonian can be presented in the form
\begin{gather}\label{SysHam}
	\hat H_{\rm{S}} = \hat H_{\rm{S1}} + \hat H_{\rm{S2}} + \hat V
\end{gather}
When interaction between subsystem is small compared to relaxation rates, then the Lindblad relaxation operators are transition operators between the eigenstates of the Hamiltonian $H_{\rm{S1}} + H_{\rm{S2}}$ of non-interacting subsystems.
When interaction between subsystems is large compared to relaxation rates and, as a consequence, splitting between eigenstates is larger than the relaxation rates, then the Lindblad superoperators are transition operators between the eigenstates of the Hamiltonian $H_{\rm{S}}$ of interacting subsystems~\cite{kosloff2013quantum,LOC2}
In the intermediate case, when the splitting between the eigenstates are comparable with the relaxation rates, one can use PS approach \cite{PS1, PS2} which takes into account the terms in the master equation which oscillate at the Rabi interaction frequency.
Moreover, PS approach results in cross-relaxation processes, when the state of one subsystems influences on relaxation of another subsystems \cite{PS4}.

From the master equation for the system density matrix~(\ref{TotHam}), one can obtain the equation for the expected values of any system operators $\hat A$ as $d\langle \hat A \rangle / dt = {\rm{Tr}}\left(\hat A  \partial \hat \rho/ \partial t\right)$.
For example, if one consider two coupled oscillators or TLSs and the amplitudes of their excitation, then the equations describing amplitude' dynamics, obtained in the local approach, predict the existence of an EP \cite{Miri_Alu, PS4, LG}.
At the same time, the global approach does not predict the EP found in local approach.
PS approach reproduces only the signature of the EP \cite{PS2, PS4, LG}.
The last means that instead of the point of eigenvalue' coincidence there is an inflection point in the dependence of the system eigenvalues on the coupling constant.

Our goal is to investigate the behavior of energy flow in the vicinity of the EP.
In the case of independent reservoirs, the Lindblad superoperators $\hat L_j$ describe the relaxation to each reservoir independently.
Thus, it is possible to determine the energy flow from the system to each reservoir.
Total energy flow from the system can be defined as follows
\begin{equation}\label{EnFlow}
\dot{H}_S=\frac{d}{dt}\langle\rho\hat H_S\rangle=\langle\dot{\rho}\hat H_S\rangle=\sum_j J_j.
\end{equation}
Here $J_j$  is energy flow in the $j$-th reservoir.

To be specific, we consider the system of two coupled TLSs in dipole approximation. 
Then, ${\hat H_S}$ in Eq.~(\ref{TotHam}) is the Hamiltonian of two coupled TLSs in the rotating-wave approximaiton without diamagnetic terms~\cite{agarwal1974quantum}:
\begin{equation}\label{H_S}
{\hat H_S} = \omega {}_1\hat \sigma _1^\dag {\hat \sigma _1} + {\omega _2}\hat \sigma _2^\dag {\hat \sigma _2} + \Omega \left( {\hat \sigma _1^\dag {{\hat \sigma }_2} + \hat \sigma _2^\dag {{\hat \sigma }_1}} \right).
\end{equation}
Here $\omega_{1,2}$ are eigenfrequencies of TLSs, $\Omega$ is the Rabi interaction constant between them, $\hat \sigma_{1,2}$ are transition operator from the excited state $|e\rangle_{1,2}$ to the ground state $|g\rangle_{1,2}$ of the first and second TLS, respectively.
Such type of Hamiltonian is valid in the case when the Rabi interaction constant $\Omega \le 0.1\omega_{1,2}$~\cite{frisk2019ultrastrong} which we will use below.
The ${\hat H_R}$ in Eq.~(\ref{TotHam}) represents the Hamiltonian of two bosonic reservoirs. 
\begin{equation}
{\hat H_R} = \sum\limits_k {\hbar {\omega _k}\hat e_{1k}^\dag {{\hat e}_{1k}}}  + \sum\limits_m {\hbar {\omega _m}\hat e_{2m}^\dag {{\hat e}_{2m}}}. 
\end{equation}
Here operators $\hat e_{1k}$, $\hat e_{2m}$ denote annihilation operators of $k$-th and $m$-th modes of the first and second reservoir, respectively, $\omega_k$ and $\omega_m$ are reservoir' eigenfrequencies.
The ${\hat H_{SR}}$ in Eq.~(\ref{TotHam}) represents the Hamiltonian of the interaction between TLSs and their reservoirs. 
We suppose that each of TLSs interacts with own bosonic reservoir as follows:
\begin{align}
\
&{\hat H_{SR}} = \epsilon \Big( \sum\limits_k {{\chi _{1k}}} (\hat \sigma_1^ +  + {{\hat \sigma}_1})(\hat e_{1k}^ +  + {{\hat e}_{1k}}) +\\
& \sum\limits_m {{\chi _{2m}}} (\hat \sigma_2^ +  + {{\hat \sigma}_2})(\hat e_{2m}^ +  + {{\hat e}_{2m}}) \Big) = \epsilon {\hat S_1}{\hat R_1} + \epsilon {\hat S_2}{\hat R_2}. \nonumber
\end{align}
Here ${\chi _{1k}},{\chi _{1m}}$ denote the coupling strengths of $k$-th and $m$-th modes of the first and second reservoir with first and second TLSs, respectively.

\section{Energy flows in the local and global approaches}
First, we investigate the behavior of the energy flow in the vicinity of the transition through the EP in the local approach and obtain explicit expression for energy flow dependence on the coupling constant.
In the local approach, the master equation for two coupled TLS is \cite{rivas2012time, LOCvsGLB2}
\begin{align}\label{p_LOC}
&\frac{{\partial \hat \rho }}{{\partial t}} =  - i\left[ {{{\hat H}_S},\hat \rho } \right] + \\ \nonumber
&+ \frac{{{G_1}( - {\omega _1})}}{2}\hat L[\hat\sigma_1,\hat\sigma_1^\dag] +\frac{{{G_1}({\omega _1})}}{2}\hat L[\hat\sigma_1^\dag,\hat\sigma_1] + \\ \nonumber
& + \frac{{{G_2}( - {\omega _2})}}{2}\hat L[\hat\sigma_2,\hat\sigma_2^\dag]+ \frac{{{G_2}({\omega _2})}}{2}\hat L[\hat\sigma_2^\dag,\hat\sigma_2].
\end{align}
Here ${G_j}(\pm\omega)=\gamma_j(\omega)(n_j(\omega)+1/2\mp1/2)$  is Fourier transformation of correlation function of $j$-th reservoir, $\gamma_j\left(\omega\right) = \pi D_j\left(\omega\right) |\chi_j\left(\omega\right)|^2$, $D_j\left(\omega\right)$ is the $j$-th reservoir density of states, $j=1,2$, \cite{PS2}.
The Lindblad superoperator has the form $\hat L[\hat X,\hat Y,\hat \rho] \equiv \hat L [\hat X, \hat Y] = 2\hat X \hat \rho \hat Y-\hat Y \hat X \hat\rho -\hat\rho \hat Y \hat X$.
Using the master equation of the local approach, Eq.~(\ref{p_LOC}), one can get the system of equations for TLS's occupancies which has the form
\begin{equation}\label{SSeq_LOC}
\frac{{d\overrightarrow{\left\langle {{{ \sigma }^\dag } \sigma } \right\rangle} }}{{dt}} = {M_{\rm{L} }}\overrightarrow{\left\langle {{{ \sigma }^\dag } \sigma } \right\rangle}  + {\overrightarrow{G}_{\rm{L} }},
\end{equation}

\begin{equation}\label{SEQ_LOC}
{M_{\rm{L}}} = \left( {\begin{array}{*{20}{c}}
{ - 2{g_1}}&0&{ - i\Omega }&{i\Omega }\\
0&{ - 2{g_2}}&{i\Omega }&{ - i\Omega }\\
{ - i\Omega }&{i\Omega }&{ - {g_1} - {g_2}+i\Delta\omega}&0\\
{i\Omega }&{ - i\Omega }&0&{ - {g_1} - {g_2-i\Delta\omega}}
\end{array}} \right),
\end{equation}

\begin{equation}
\overrightarrow{\left\langle {{{\sigma }^\dag } \sigma } \right\rangle}  = \left( {\begin{array}{*{20}{c}}
{\langle {\hat \sigma _1^\dag {{\hat \sigma }_1}} \rangle }\\
{\langle {\hat \sigma _2^\dag {{\hat \sigma }_2}} \rangle }\\
{\langle {\hat \sigma _1^\dag {{\hat \sigma }_2}} \rangle }\\
{\langle {\hat \sigma _2^\dag {{\hat \sigma }_1}} \rangle }
\end{array}} \right),\ \ \ \ 
{\overrightarrow{G}_{L }} = \left( {\begin{array}{*{20}{c}}
{{G_1}({\omega _1})}\\
{{G_2}({\omega _2})}\\
0\\
0
\end{array}} \right).
\end{equation}
Here ${g_j} = ({{G_j}( - {\omega _j}) + {G_j}({\omega _j})})/2=\gamma_j(\omega_j)(n_j(\omega_j)+1/2)$ are relaxation rates, $n_j(\omega)=1/(\exp(\omega/T_j)-1)$ is the mean value of quanta in $j$-th reservoir at freaquency $\omega$ at temperature $T_j$, $\Delta\omega=\omega_1-\omega_2$ is the detuning between TLSs.

\begin{figure}[h]
\begin{minipage}[h]{0.95\linewidth}
\centering{\includegraphics[width=1\linewidth]{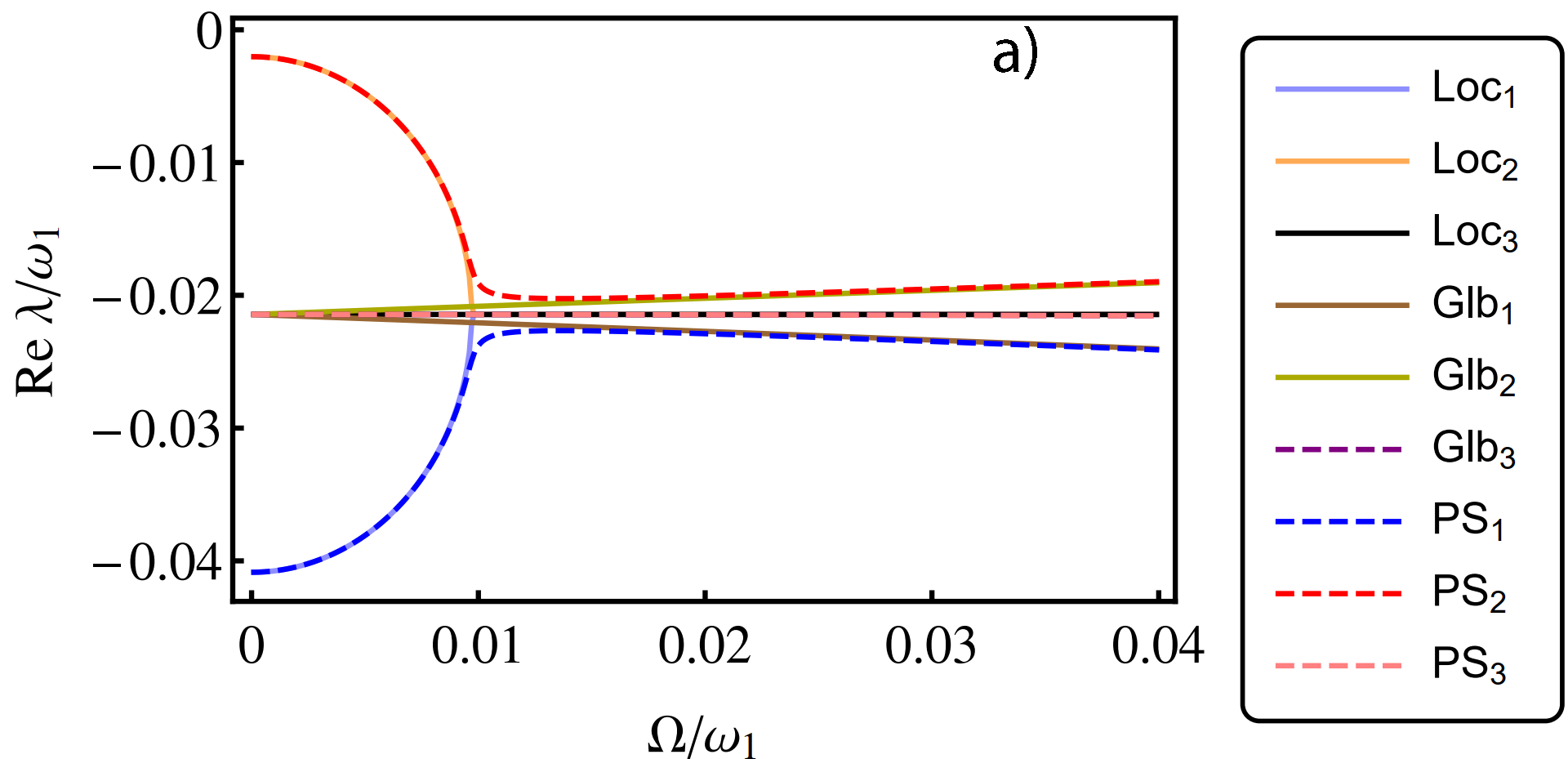}}
\end{minipage}
\vspace{10pt}

\begin{minipage}[h]{0.95\linewidth}
\centering{\includegraphics[width=1\linewidth]{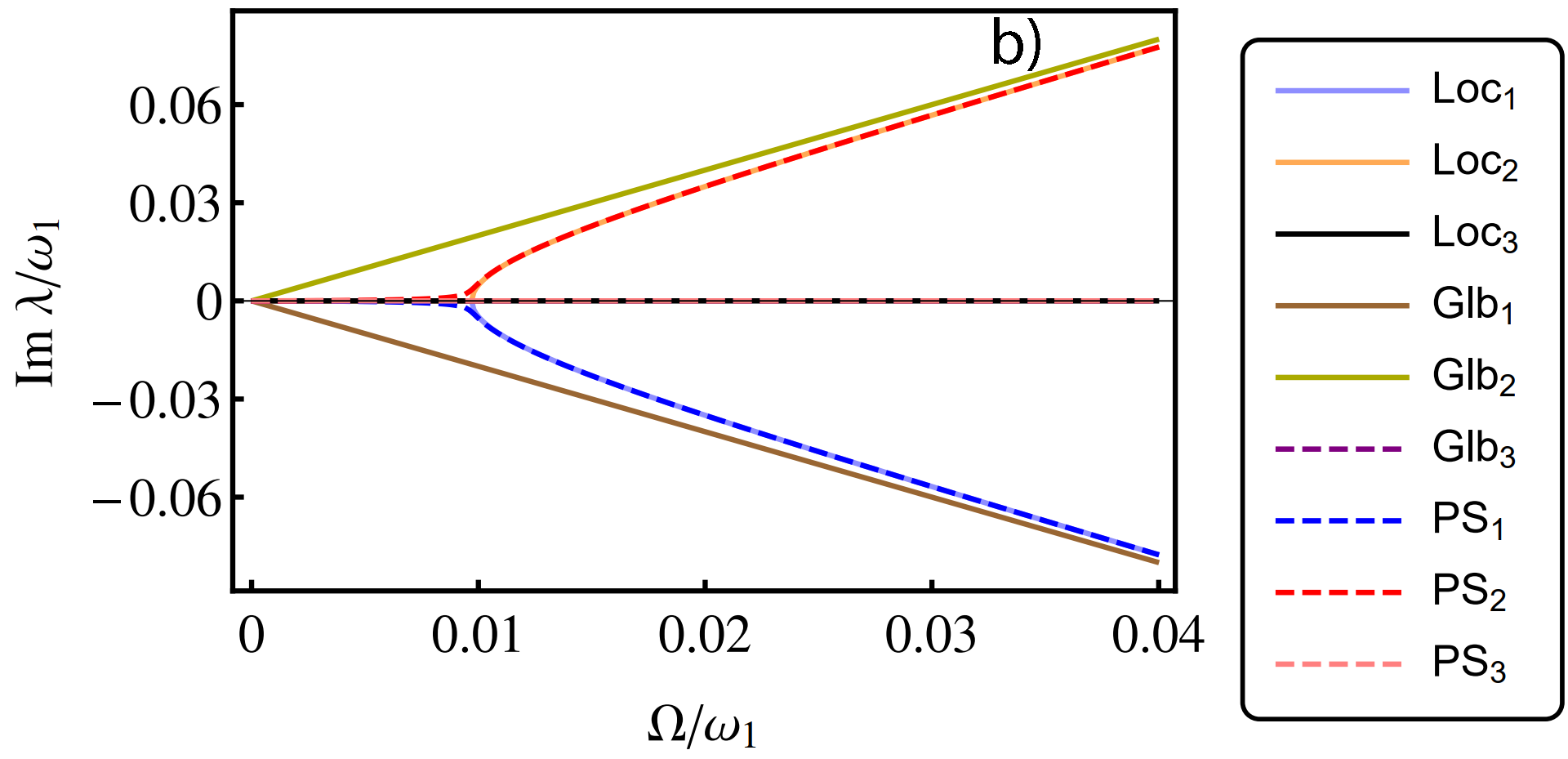}}
\end{minipage}
	\caption{Real a) and imaginary b) parts of eigenvalues $\lambda$ of matrices $M_L$, $M_G$, $M_{PS}$ from Eqs. (\ref{SSeq_LOC}), (\ref{SSeq_GLB}), (\ref{SSeq_PS}).  The case $\gamma_k(\omega)=c_k\omega^3$,  $c_1=0.002$, $c_2=0.04$, $T_1=0.2$, $T_2=0.22$, $\Delta \omega=0$, is depicted. The forth eigenvalue coincide with the third one in all approaches.}
\label{roots_split}
\end{figure}

If $\Delta \omega = 0$, the eigenvalues of $M_{{\rm{L}}}$, shown in Fig.\ref{roots_split}, have the form (see also Appendix A)
\begin{equation}
\lambda_{1,2} = -(g_1+g_2),
\end{equation}
\begin{equation}
\lambda_\pm=-(g_1+g_2)\pm\sqrt{(g_1-g_2)^2+4\Omega^2}.
\end{equation}
There is an EP when $\left|g_1 - g_2\right| = 2\Omega$.
Using Eqs.~(\ref{LindGen}), (\ref{EnFlow}), (\ref{H_S}), and (\ref{SSeq_LOC}) one can obtain the following expression for the energy flow $J_1$ from the system to the first reservoir:
\begin{gather}\label{J1}
J_1=- 2{\omega _1}{g_1}\langle {\hat \sigma _1^\dag {{\hat \sigma }_1}} \rangle  + {\omega _1}{G_1}({\omega _1}) - 2\Omega {g_1}{\mathop{\rm Re}\nolimits} {\langle {\hat \sigma _1^\dag {{\hat \sigma }_2}} \rangle} .
\end{gather}
Substituting the expressions for $\langle \hat \sigma^{\dag}\hat \sigma_1 \rangle$ and $\langle \hat \sigma^{\dag}_1 \hat \sigma_2 \rangle$ from the stationary solution of Eq.~(\ref{SSeq_LOC}), we obtain (see Appendix B)
\begin{gather}\label{Jstt}
J_1^{st}=  - \frac{{\left( {{\omega _1}{g_2} + {\omega _2}{g_1}} \right){\gamma _1}({\omega _1}){\gamma _2}({\omega _2})}\Omega^2}{{2{{\left( {{g_1} + {g_2}} \right)}^2}}\left({g{}_1g{}_2 + \frac{{{{(\Delta \omega )}^2}}}{{{{\left( {g{}_1 + {g_2}} \right)}^2}}}g{}_1g{}_2 + {\Omega ^2}}\right)}\times
	\\ \nonumber
\times\frac{{{e^{{\omega _1}/{T_1}}} - {e^{{\omega _2}/{T_2}}}}}{{\left( {{e^{{\omega _1}/{T_1}}} - 1} \right)\left( {{e^{{\omega _2}/{T_2}}} - 1} \right)}}.
\end{gather}

The absolute value of stationary energy flow $J_1^{st}$ monotonically grows with the increase of $\Omega$.
For small $\Omega$, $J_1^{st}\sim \Omega^2$, while for large $\Omega$,  $J_1^{st}\simeq \mathrm{const}$.
The specific energy flow is defined as
\begin{gather}\label{j_spe}
	j_1^{st} = J_1^{st}/\Omega,
\end{gather}

From Eq.~(\ref{Jstt}) one can evaluate the value of $\Omega$ at which absolute value of $j_1^{st}$ is maximized.
It is determined by the following expression (see Appendix B):
\begin{equation}\label{Wip}
\Omega_{max}=\sqrt {g_1 g_2\left( {1 + \frac{{{{(\Delta \omega )}^2}}}{{{{\left( {g_1 + {g_2}} \right)}^2}}}} \right)}.
\end{equation}

\begin{figure}
\begin{minipage}{0.85\linewidth}
\centering{\includegraphics[width=\linewidth]{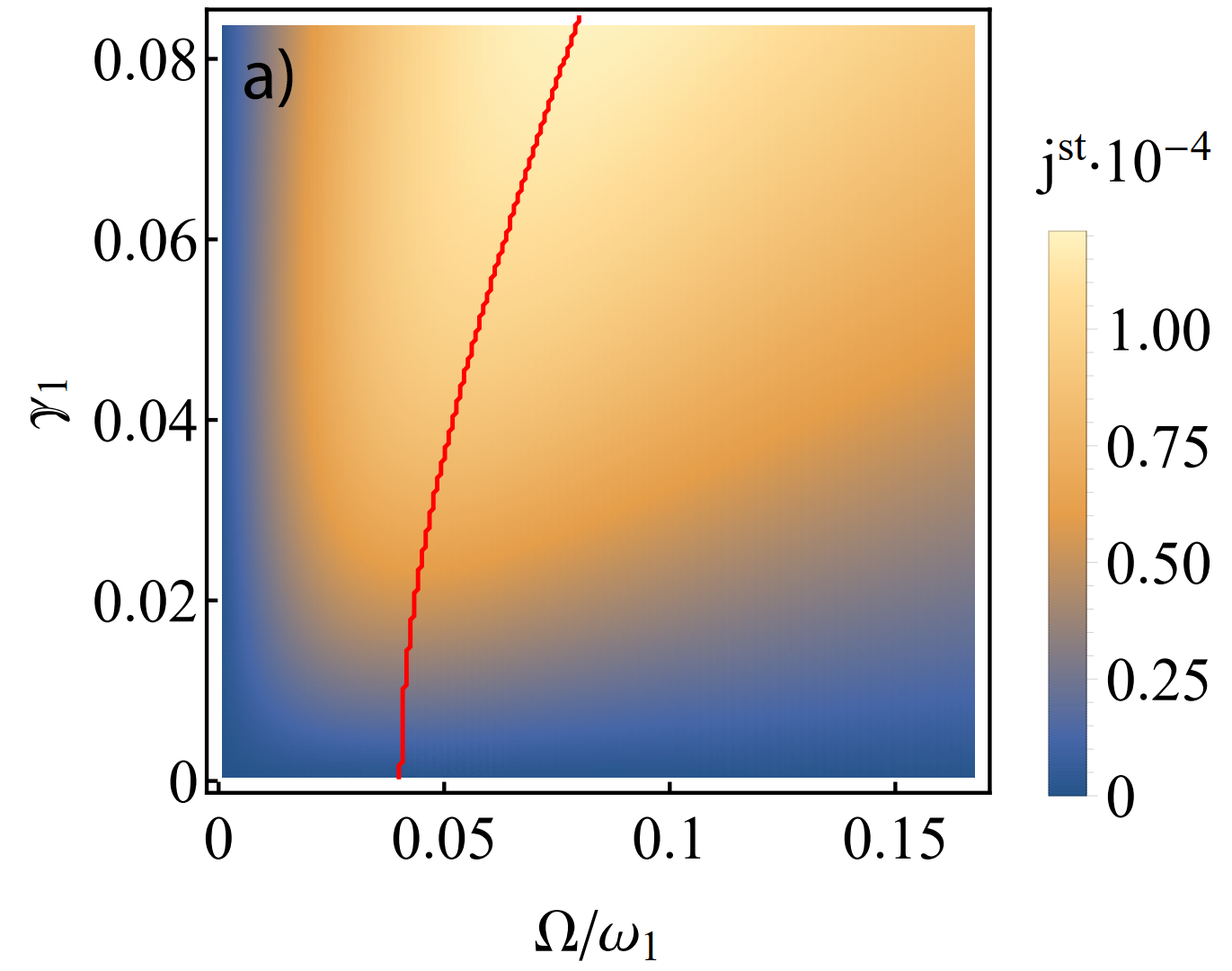}}
\end{minipage}
\vspace{10pt}

\begin{minipage}{0.85\linewidth}
\centering{\includegraphics[width=\linewidth]{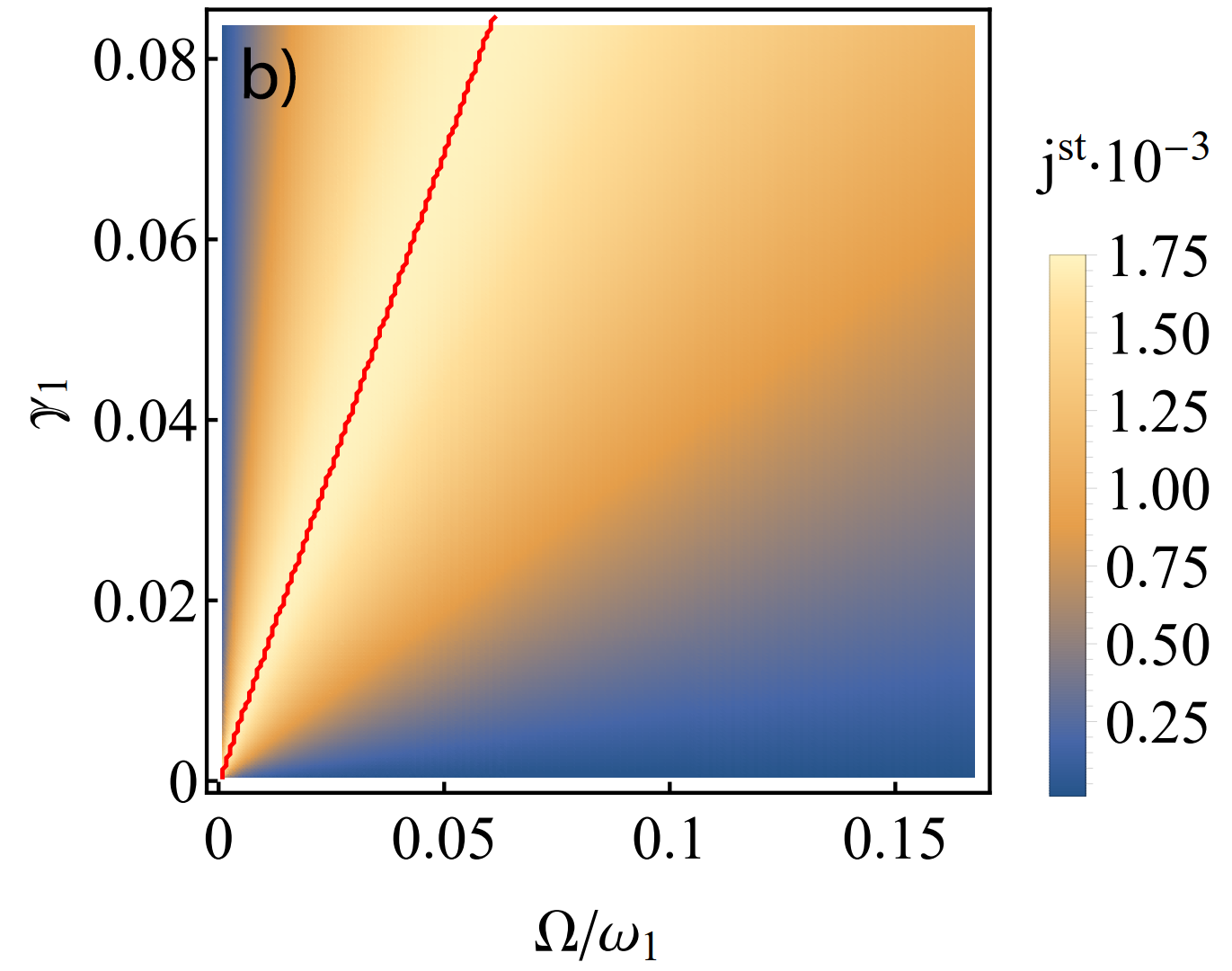}}
\end{minipage}
\caption{Stationary specific energy flow $j^{st}$ to the hot reservoir (see Eq. (\ref{j_spe})) in local approach as function of $\gamma_1$ and $\Omega$, $\omega_1=1.0$, $\gamma_k(\omega)=c_k\omega^3$,  $c_1=c_2/2$, $T_1=0.2 \omega_1$, $T_2=0.22 \omega_1$, a) $\Delta \omega=-0.09\omega_1$; b) $\Delta \omega=0$. Red lines depict maximum of $j^{st}$ over $\Omega$ for fixed $\gamma_1$.}
\label{LOC_2D}
\end{figure}

In the Fig. \ref{LOC_2D}, the specific energy flow $j^{st} = -j_1^{st}=j_2^{st}$ as a function of $\Omega$ and $\gamma_1$ is shown in the case $\gamma_1=\gamma_2/2$.
It is seen that at any fixed value of $\gamma_1$ there is a value of $\Omega_{{\rm{max}}}$ at which $j^{st}$ reaches maximal value. 
The function $\Omega_{max}=\Omega_{max}\left(\gamma_1\right)$, as the value of $\Omega$ at which $j^{st}$ is maximized is shown by the red line in Fig.~\ref{LOC_2D}.
In the case of zero detuning, this line starts from zero values of $\Omega$ and $\gamma_1$.
However, in the case of non-zero detuning this line starts from zero value of $\gamma_1$ but non-zero value of $\Omega$.
The last means that there exists a critical value of $\Omega_{cr}\simeq|\Delta\omega|$, at which specific energy flow reaches its maximal value.
Thus, at non-zero detuning, to specific energy flow be maximized, it is necessary to exceed threshold value of $\Omega=\Omega_{cr}$.

On the other hand,  if $\gamma_1=c\gamma_2$ ($c$ is arbitrary constant) and $\Omega$ is fixed, $J^{st}\sim \gamma_2/(c\gamma_2^2+c(\Delta\omega)^2/(c+1)^2+\Omega^2)$.
Thus, $J^{st}$ grows as $J^{st}\sim \gamma_2$, when $\gamma_2\ll|\Delta\omega|+\Omega$, and is suppressed as $J^{st}\sim 1/\gamma_2$, when $\gamma_2\gg|\Delta\omega|+\Omega$.
From this it follows that stationary energy flow has maximum with respect to $\gamma_2$.
Such behaviour is seen in Fig. \ref{LOC_2D}, when for fixed $\Omega$ stationary specific energy flow $j^{st}$ (as well as $J^{st}=j^{st}\Omega$) is suppressed by high values of $\gamma_1$ (also see Fig. \ref{LOC_PS_J} in Appendix B).

The local approach has one significant drawback.
Namely, it predicts violation of the second law of thermodynamics.
Indeed, from the Eq. (\ref{Jstt}), it is seen, that  $\mathrm{sign}(J_1^{st})=-\mathrm{sign} (e^{{\omega_1}/{T_1}} - e^{{\omega_2}/{T_2}})$.
If two conditions, $T_1>T_2$ and $\frac{{{\omega _1}}}{{{T_1}}} > \frac{{{\omega _2}}}{{{T_2}}}$, are fulfilled, then $J_1^{st}<0$, i.e., the energy flows from the system to the hot reservoir.
Thus, the second law of thermodynamics is violated.
Note that this behavior also takes place for coupled oscillators~\cite{LOC1}.

As has been mentioned in the Introduction, second law of thermodynamics is restored in the global approach in which Lindblad relaxation superoperators describe transition between eigenstates of the coupled subsystems.
In the case of two coupled TLSs, the master equation has the form~\cite{DAE} 
\begin{align}\label{p_GLB}
&\frac{{\partial {{\hat { \rho} }}}}{{\partial t}} =  - i[{\hat{H}_s},{\hat{\rho} }] + \\ \nonumber
&\frac{{{G_1}( - (\tilde \omega  + \tilde \Omega ))}}{2}L[\hat A_1,\hat A_1^\dag] + \frac{{{G_1}(\tilde \omega  + \tilde \Omega )}}{2}L[\hat A_1^\dag,\hat A_1] + \\ \nonumber
&\frac{{{G_1}( - (\tilde \omega  - \tilde \Omega ))}}{2}L[\hat B_1,\hat B_1^\dag] + \frac{{{G_1}(\tilde \omega  - \tilde \Omega )}}{2}L[\hat B_1^\dag,\hat B_1] + \\ \nonumber
&\left( {(1) \mathbin{\lower.3ex\hbox{$\buildrel\textstyle\rightarrow\over{\smash{\leftarrow}\vphantom{_{\vbox to.5ex{\vss}}}}$}} (2)} \right) .  \nonumber
\end{align}
Here ${(1) \mathbin{\lower.3ex\hbox{$\buildrel\textstyle\rightarrow\over{\smash{\leftarrow}\vphantom{_{\vbox to.5ex{\vss}}}}$}} (2)}$ denotes the same dissipative terms where index $1$ is exchanged with index $2$, $\tilde \omega  = ({\omega _1} + {\omega _2})/2$, $\tilde \Omega  = \sqrt {{\Delta ^2} + 4{\Omega ^2}} /2$.
The dissipative operators have the form
\begin{align}
& {\hat A_1} =\frac{1}{2r} \left( {{\hat \sigma _1}\frac{{y + 2r }}{2} + {\hat \sigma _2} - 2\hat{\sigma} _1^\dag {\hat \sigma _1}{\hat \sigma _2}} \right) \\ \nonumber
& {\hat B_1} =\frac{1}{2r} \left( { - {\hat \sigma _1}\frac{{y - 2r }}{2} - {\hat \sigma _2} + 2\hat {\sigma}_1^\dag {\hat \sigma _1}{\hat \sigma _2}} \right) \\ \nonumber
& {\hat A_2} =\frac{1}{2r} \left( {{\hat \sigma _2}\frac{{ - y + 2r }}{2} + {\hat \sigma _1} - 2\hat {\sigma}_2^\dag {\hat \sigma _2}{\hat \sigma _1}} \right) \\ \nonumber
& {\hat B_2} =\frac{1}{2r} \left( {{\hat \sigma _2}\frac{{y + 2r }}{2} - {\hat \sigma _1} + 2\hat {\sigma}_2^\dag {\hat \sigma _2}{\hat \sigma _1}} \right)	\nonumber
\end{align}
where $r=\sqrt {{y^2}/4 + 1} $, $y = \left( {{\omega _1} - {\omega _2}} \right)/\Omega$.
We neglect in this equation the terms that describe the shift of the mode frequency and do not affect the relaxation \cite{B-P}.
From Eq. (\ref{p_GLB}), one can obtain the following system for the TLSs occupancies and energy flow:
\begin{equation}\label{SSeq_GLB}
\frac{{d\overrightarrow{\left\langle {{{ \sigma }^\dag } \sigma } \right\rangle} }}{{dt}} = {M_{\rm{G} }}\overrightarrow{\left\langle {{{ \sigma }^\dag } \sigma } \right\rangle}  + {\overrightarrow{G}_{\rm{G} }},\ \ \ \ 
{\overrightarrow{G}_{G}} = \left( {\begin{array}{*{20}{c}}
{{S_{11,G}}}\\
{{S_{22,G}}}\\
{{S_{12,G}}}\\
{{S^*_{12,G}}}
\end{array}} \right),
\end{equation}
\begin{align}\label{SEQ_GLB}
&{M_{\rm{G}}} =\\ \nonumber
&=\left( {\begin{array}{*{20}{c}}
A_G&0&{ - i\Omega  + Q_G}&{i\Omega  + Q^*_G }\\
0&F_G&{i\Omega  + T_G}&{ - i\Omega  + T^*_G }\\
{ - i\Omega  + T^*_G }&{i\Omega  + Q^*_G }&{i\Delta \omega  + Y_G}&0\\
{i\Omega  +T_G}&{ - i\Omega  + Q_G}&0&{ - i\Delta \omega  +Y^*_G}
\end{array}} \right).
\end{align}
(see Appendix C for explicit expressions for coefficients)
Imaginary parts of eigenvalues of the matrix $M_{{\rm{G}}}$ are always split (see Fig. \ref{roots_split}).
Thus, the system~(\ref{SSeq_GLB}) does not predict an EP.
The consequence of this fact is that the value of $\Omega$ at which specific energy flow $j^{st}$ achieves maximum $j^{st}_{max}$ does not depend on $\gamma$, see Fig.2. 

The dependence $J^{st}(\Omega)$ in global approach differs from that obtained in local approach (see Appendix D).
If $\Delta\omega\ne0$, it first reaches the maximum and after that decreases and tends to the value given by PS approach.
If $\Delta\omega=0$, it starts from the maximal value and decreases.
\begin{figure}
\begin{minipage}{0.85\linewidth}
\includegraphics[width=\linewidth]{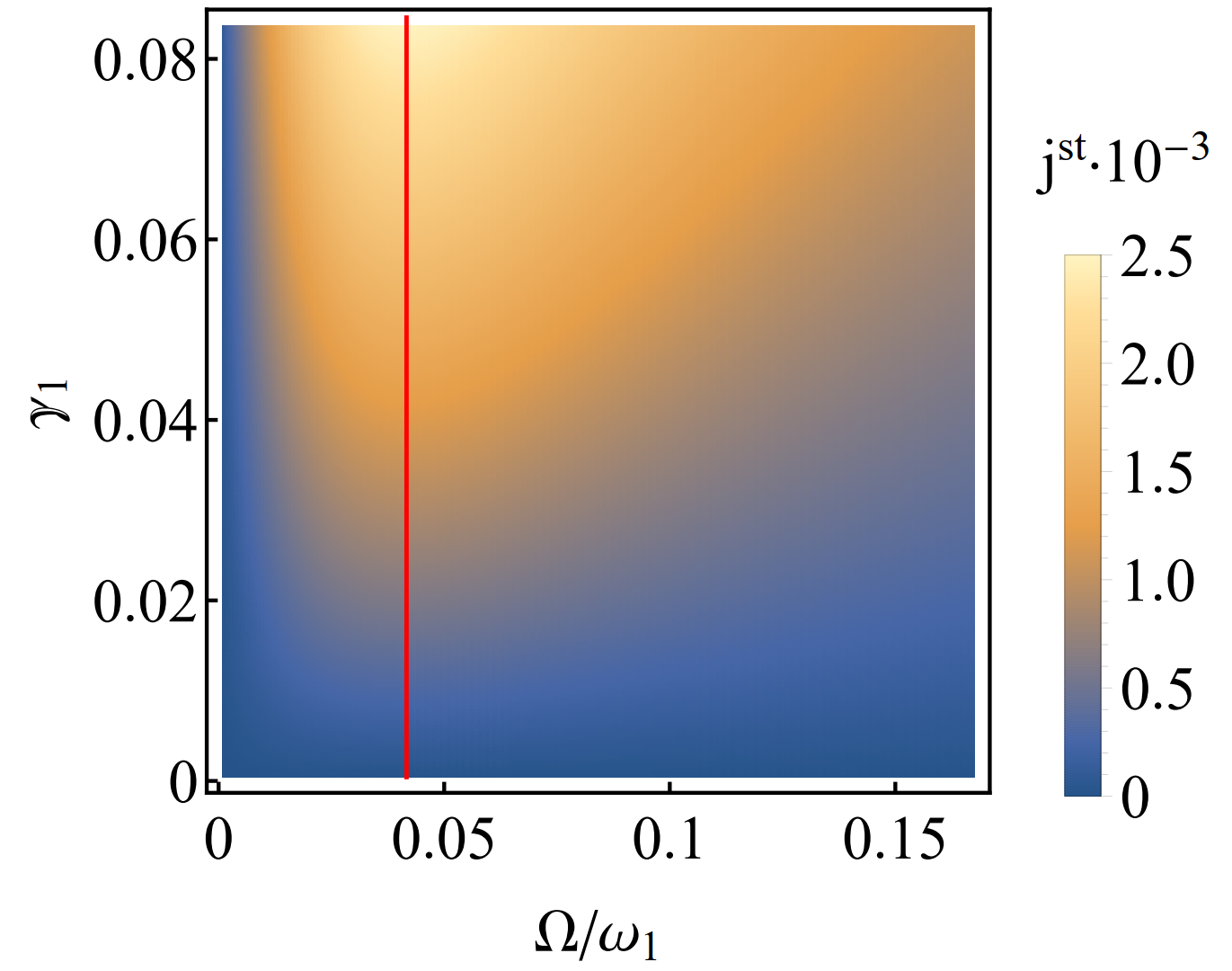}
\end{minipage}
	\caption{Stationary specific energy flow $j^{st}$ to the hot reservoir in global approach as function of $\gamma_1$ and $\Omega$, $\omega_1=1.0$, $\gamma_k(\omega)=c_k\omega^3$,  $c_1=c_2/2$, $T_1=0.2\omega_1$, $T_2=0.22\omega_1$, $\Delta \omega=-0.09\omega_1$. Red lines depict maximum of $j^{st}$ over $\Omega$ for fixed $\gamma_1$.}
\end{figure}
Contrary to the local approach, stationary energy flow for fixed $\Omega$ is not suppressed by high values of $\gamma_1$.
Moreover, it linearly grows along with $\gamma_1$ (see Appendix D).

Note that local approach is valid when $\Omega \ll \gamma$ while global approach is valid when $\Omega \gg \gamma$.
Thus, they are not applicable near the EP.
As has been shown in \cite{PS2,PS4},  one can use PS approach which is valid near the EP and is reproduced local and global approaches when $\Omega \ll \gamma$ and $\Omega \gg \gamma$, respectively.

\section{ENERGY FLOW IN PARTIAL-SECULAR APPROACH}
In the case of two coupled TLSs, the PS approach results in the following master equation for the density matrix:
\begin{align}\label{p_PS}
&\frac{{\partial {{\hat \rho }}}}{{\partial t}} =  - i[{\hat{H}_S},{\hat{\rho} }] + \\ \nonumber
&\frac{{{G_1}( - (\tilde \omega  + \tilde \Omega ))}}{2}\hat L[\hat A_1,\hat A_1^\dag] + \frac{{{G_1}(\tilde \omega  + \tilde \Omega )}}{2}\hat L[\hat A_1^\dag,\hat A_1] + \\ \nonumber
&\frac{{{G_1}( - (\tilde \omega  - \tilde \Omega ))}}{2} \hat L[\hat B_1,\hat B_1^\dag] +  \frac{{{G_1}(\tilde \omega  - \tilde \Omega )}}{2}\hat L[\hat B_1^\dag,\hat B_1] + \\ \nonumber
&\frac{{{G_{1 - }}( - (\tilde \omega  - \tilde \Omega )) + {G_{1 + }}( - (\tilde \omega  + \tilde \Omega ))}}{2} \hat L[\hat A_1,\hat B_1^\dag] + \\ \nonumber
&\frac{{{G_{1 - }}( - (\tilde \omega  - \tilde \Omega )) - {G_{1 + }}( - (\tilde \omega  + \tilde \Omega ))}}{2}\left[ {\hat B_1^\dag {{\hat A}_1},{\hat{\rho} }} \right] + \\ \nonumber
&\frac{{{G_{1 - }}(\tilde \omega  + \tilde \Omega ) + {G_{1 + }}(\tilde \omega  - \tilde \Omega )}}{2} \hat L[\hat B_1^\dag,\hat A_1] + \\ \nonumber
&\frac{{{G_{1 - }}(\tilde \omega  + \tilde \Omega ) - {G_{1 + }}(\tilde \omega  - \tilde \Omega )}}{2}\left[ {{{\hat A}_1}\hat B_1^\dag ,{\hat{\rho} }} \right] + \\ \nonumber
&\frac{{{G_{1 - }}(\tilde \omega  - \tilde \Omega ) + {G_{1 + }}(\tilde \omega  + \tilde \Omega )}}{2}\hat L[\hat A_1^\dag,\hat B_1]+ \\ \nonumber
&\frac{{{G_{1 - }}(\tilde \omega  - \tilde \Omega ) - {G_{1 + }}(\tilde \omega  + \tilde \Omega )}}{2}\left[ {{{\hat B}_1}\hat A_1^\dag ,{\hat{\rho} }} \right] + \\ \nonumber
&\frac{{{G_{1 - }}( - (\tilde \omega  + \tilde \Omega )) + {G_{1 + }}( - (\tilde \omega  - \tilde \Omega ))}}{2}\hat L[\hat B_1,\hat A_1^\dag] + \\ \nonumber
&\frac{{{G_{1 - }}( - (\tilde \omega  + \tilde \Omega )) - {G_{1 + }}( - (\tilde \omega  - \tilde \Omega ))}}{2}\left[ {\hat A_1^\dag {{\hat B}_1},{\hat{\rho} }} \right] + \\ \nonumber
&\left( {(1) \mathbin{\lower.3ex\hbox{$\buildrel\textstyle\rightarrow\over 
{\smash{\leftarrow}\vphantom{_{\vbox to.5ex{\vss}}}}$}} (2)} \right).  \nonumber
\end{align}
Similar to Eq.~(\ref{p_GLB}), we neglect terms that just shift the mode frequency and do not affect the relaxation.
There are three types of terms in this master equation: conventional dissipative terms (first two lines in Eq.~(\ref{p_PS})), dissipative PS terms (other terms with $\hat L[\hat X,\hat Y]$) and terms with commutators.

To derive Eq. (\ref{p_PS}) one can use interaction picture for $\hat \sigma_1$ and $\hat \sigma_2$ (see Appendix in Ref.~\cite{DAE}) and the PS master equation for coupled oscillators (see Supplement Materials in Ref.~\cite{PS2} and Appendix A in Ref.~\cite{PS4}) with following substitutions $\omega\rightarrow\tilde\omega$, $\Omega\rightarrow\tilde\Omega$, $(\hat a_{1,2}^\dag+\hat a_{1,2})/\sqrt{2}\rightarrow \hat A_{1,2}$, $(\hat a_{1,2}^\dag-\hat a_{1,2})\sqrt{2}\rightarrow \hat B_{1,2}$~\cite{PS2,PS4}.

From Eq. (\ref{p_PS}), one can obtain the following system for the TLSs occupancies and energy flow
\begin{equation}\label{SSeq_PS}
\frac{{d\overrightarrow{\left\langle {{{ \sigma }^\dag } \sigma } \right\rangle} }}{{dt}} = {M_{\rm{PS} }}\overrightarrow{\left\langle {{{ \sigma }^\dag } \sigma } \right\rangle}  + {\overrightarrow{G}_{\rm{PS} }},\ \ \ \ 
{\overrightarrow{G}_{PS }} = \left( {\begin{array}{*{20}{c}}
{{S_{11}}}\\
{{S_{22}}}\\
{{S_{12}}}\\
{{S^*_{12}}}
\end{array}} \right),
\end{equation}
which is similar to Eqs.~(\ref{SSeq_LOC}) and (\ref{SSeq_GLB}).
Here
\begin{align}\label{SEQ_PS}
&{M_{\rm{PS}}} =\\ \nonumber
&=\left( {\begin{array}{*{20}{c}}
A&0&{ - i\Omega  + Q}&{i\Omega  + Q^* }\\
0&F&{i\Omega  + T}&{ - i\Omega  + T^* }\\
{ - i\Omega  + T^* }&{i\Omega  + Q^* }&{i\Delta \omega  + Y}&0\\
{i\Omega  +T}&{ - i\Omega  + Q}&0&{ - i\Delta \omega  +Y^*}
\end{array}} \right).
\end{align}
Here $Y=(A+F)/2+D$, $Q=B+C$, $T=B-C$. 
Coefficients  $A$, $F$ are responsible for relaxation rates of occupancies and coefficient $B$ is responsible for cross-relaxation rates.
Coefficients $C$, $D$ are formed from terms with commutators in Eq.~(\ref{p_PS}) (explicit expressions for $A$, $F$, $B$, $C$, $D$ see Appendix C.)

From Fig. \ref{roots_split} it is seen that eigenvalues of the matrix in Eq. (\ref{SEQ_PS}) are always splitted, and there is only a signature of the EP.
It is seen that the PS approach asymptotically coincides with local approach in the range $\Omega\ll\gamma_{1,2}$ and asymptotically coincides with global approach in the range $\Omega\gg\gamma_{1,2}$

\begin{figure}[h]
\begin{minipage}{0.85\linewidth}
\centering{\includegraphics[width=\linewidth]{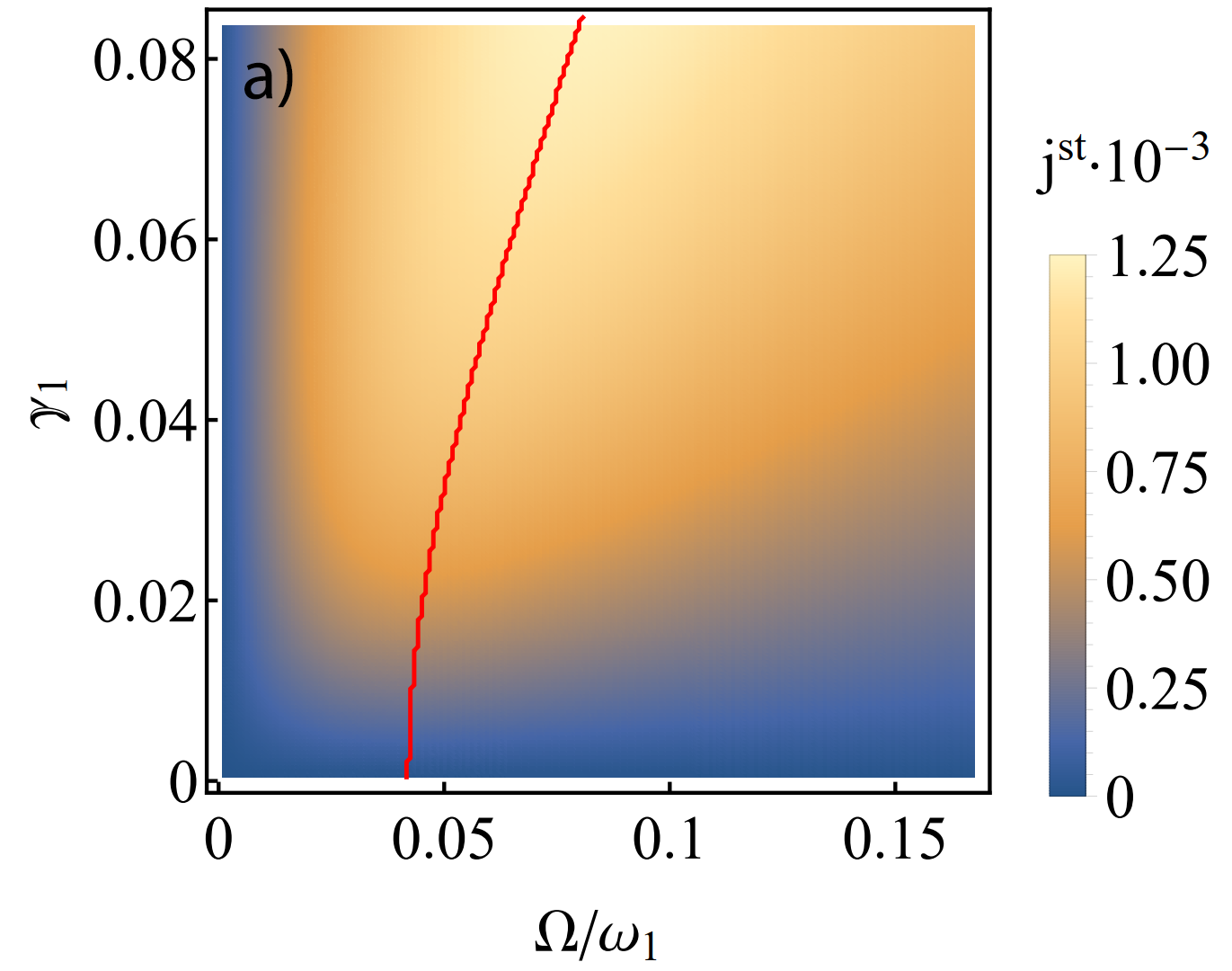}}
\end{minipage}
\vspace{10pt}

\begin{minipage}{0.85\linewidth}
\centering{\includegraphics[width=\linewidth]{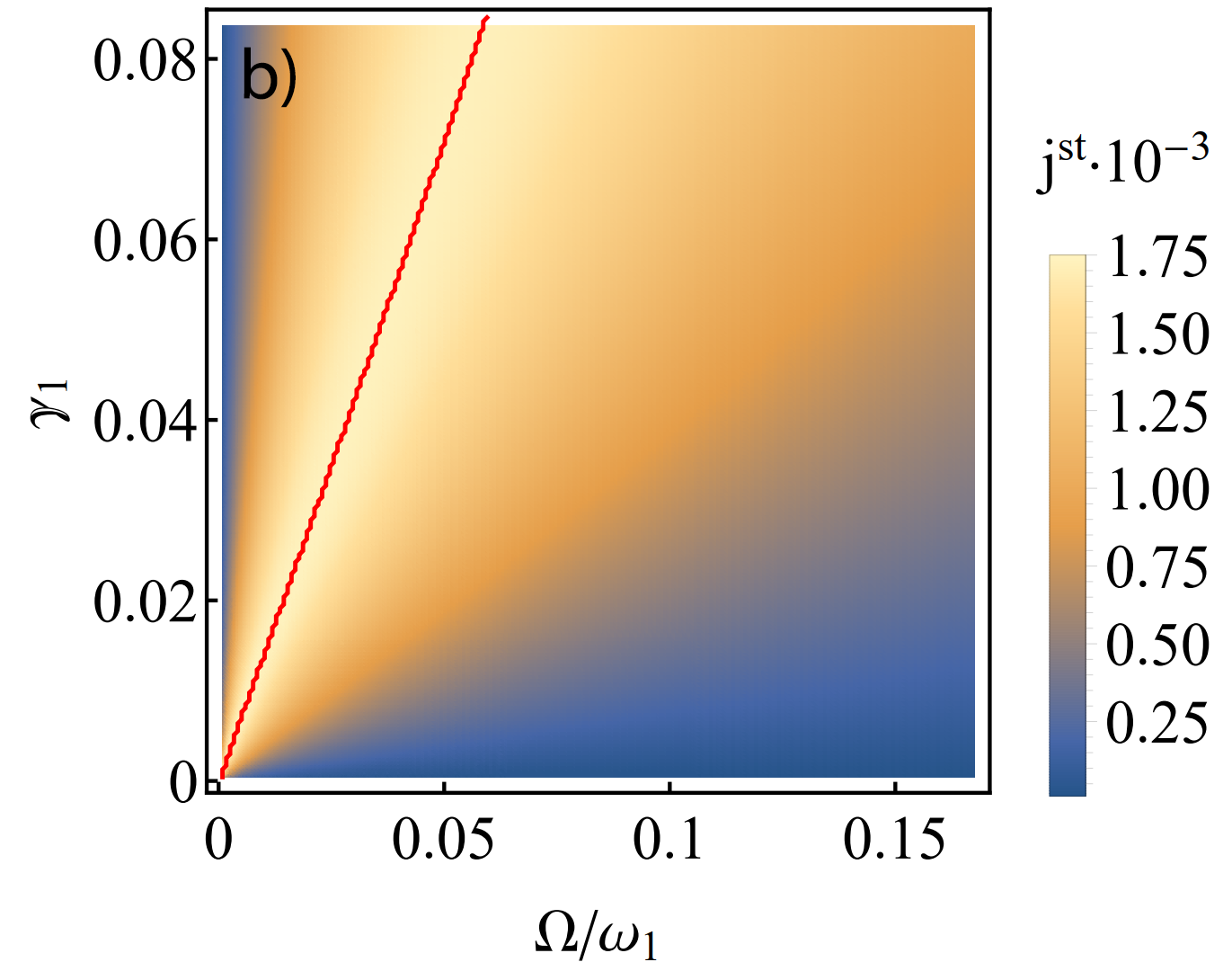}}
\end{minipage}
\caption{Stationary specific energy flow $j^{st}$ to the hot reservoir in PS approach as function of $\gamma_1$ and $\Omega$, $\omega_1=1.0$, $\gamma_k(\omega)=c_k\omega^3$,  $c_1=c_2/2$, $T_1=0.2$, $T_2=0.22$, a) $\Delta \omega=-0.09\omega_1$; b) $\Delta \omega=0$. Red lines show maximum of $j^{st}$ over $\Omega$ for fixed $\gamma_1$.}
\label{PS_2D}
\end{figure}

We numerically solve the system of equations (\ref{SSeq_PS}).
The results for the stationary specific energy flow $j^{st}$ as a function of $\Omega$ and $\gamma$ are shown in Fig. \ref{PS_2D}.
It is seen that the numerical simulation justify the local approach qualitatively.
However, second law of thermodynamics is always fulfilled, in particular, in the parameter range, where it is violated in the local approach.

In addition, the PS approach qualitatively confirms the behavior of optimal line in the case of non-zero detuning in the local approach, see Fig. \ref{PS_2D}a.
Namely, at non-zero detuning, there is critical value of $\Omega_{cr}\simeq |\Delta \omega|$, below which stationary specific energy flow $j^{st}$ does not reach its maximal value with respect to $\Omega$.
Thus, to achieve maximization of specific energy flow, it is necessary to reach a critical value of coupling constant.

The dependence $J^{st}(\Omega)$ reproduces the behaviour of the stationary energy flow in local approach when $\Omega\ll\gamma_{1,2}$ and reproduces the behaviour of stationary energy flow in global approach while $\Omega\gg\gamma_{1,2}$.  This results in the suppression of energy flow for fixed value of $\Omega$ by high values of relaxation rate $\gamma_1$ (see Fig.~\ref{PS_2D} and Fig.~\ref{LOC_PS_J}).

\section{Discussion and conclusion}
In this paper, we have shown that stationary specific energy flow between two TLSs both interacted with own reservoirs has maximum at some value  $\Omega=\Omega_{max}$ of the coupling constant between TLSs.
This result is valid in local, global and PS approaches.
In the global approach $\Omega_{max}$ does not depend on relaxation rates of TLSs.
In the local and PS approaches $\Omega_{max}$ depends on relaxation rates of TLSs.
We have shown that in the case of non-zero detuning in all three approaches there exists the non-zero critical value of $\Omega=\Omega_{cr}\sim\Delta\omega$ at which specific energy flow is maximized at an arbitrary relaxation rates.
We have demonstrated that stationary specific energy flow is suppressed in local and PS approaches at large values of relaxation rates, when $\Omega$ is fixed.
Notably, for small values of $\Omega$ the PS approach tends to the local approach, and for large values of $\Omega$ the PS approach tends to the global approach. 
It is worth noting, that energy flow from the system to reservoirs depicts their ability for energy exchange.
While stationary specific energy flow decreases for high $\Omega$, the energy flow (not divided by $\Omega$) asymptotically reaches its maximal value.
That means that relaxation rate of individual subsystems might be considered as factors which limit the energy flow.

Maximization of specific energy flow at nonzero detuning can serve as an criterion for transition to the strong coupling regime.
Indeed, in the case $\Delta\omega\gg\gamma_{1,2}$ eigenvalues are always splitted both in the local, global and PS approaches.
Thus, from this point of view, even if $\Omega=0$ the system might be considered to be in the strong coupling regime.
It obviously has no physical meaning.
In contrast, stationary specific energy flow is maximized at nonzero $\Omega$ at nonzero detuning.
In the case $\Delta\omega=0$ the system usually is considered to be in the strong coupling regime if eigenfrequencies are splitted, i.e., when $\Omega>\Omega_{EP}\sim |\gamma_1-\gamma_2|$.
The maximization of specific energy flow occurs at $\Omega=\Omega_{max}\sim\sqrt{\gamma_1\gamma_2}$.
Thus, the maximization of the specific energy flow indicates the transition to the strong coupling regime.

Finally, let us discuss the relation between considered models and more complex approaches to the description of dissipative quantum system.
Previous researches have investigated the dependence of energy flow between system and reservoir on the coupling between them~\cite{NIBA1, NIBA2, RC1, SSH1}.
It has been found that energy flow from hot reservoir to the cold one grows while coupling constant between the system and reservoirs is growing, after that experience a maximum and, finally, is suppressed at larger coupling with reservoir. 
These results are usually obtained beyond the Born-Markov approximation, e.g., in Noninteracting-blip approximation~\cite{NIBA1,NIBA2,NIBA3}, Reaction Coordinate (RC) method  \cite{RC1,RC2,RC3}, Keldysh nonequilibrium Green's function \cite{RC1}, and Stochastic surrogate Hamiltonian~\cite{SSH1,SSH3}.
In our work we have established the suppression of stationary energy flow at high values of dissipative rates in the local and the PS approaches which is in agreement with more complicated models.

\begin{acknowledgments}
The research was financially supported by a grant from Russian Science Foundation (project No. 20-72-10057).
I.V.V. thanks the Foundation for the Advancement of Theoretical Physics and Mathematics ``Basis''.
\end{acknowledgments}

\bibliography{Local_Approach}

\begin{widetext}
\appendix
\section{Derivation of  Eq. (\ref{SEQ_LOC})}
In this Appendix we find eigenvalues $\lambda$ of matrix from Eq. (\ref{SEQ_LOC}).
To do this, we equate determinant of $M_{{\rm{L}}}$ to zero:

\begin{gather}
0 = \det M_{\rm{L}} = -(2{g_1}+\lambda)\Big((2{g_2}+\lambda)( - {g_1} - {g_2}+i\Delta\omega-\lambda)( {g_1}+ {g_2}+i\Delta\omega+\lambda)-
	\\ \nonumber
- \Omega^2( {g_1} + {g_2}+i\Delta\omega+\lambda)-\Omega^2( {g_1} + {g_2}-i\Delta\omega+\lambda)\Big)-i\Omega\left((2{g_2}+\lambda)2i\Omega(g_1+g_2+\lambda)\right)=
	\\ \nonumber
= (2g_1+\lambda)\left((2g_2+\lambda)((g_1+g_2+\lambda)^2+\Delta\omega^2)+2\Omega^2(g_1+g_2+\lambda)\right)+2\Omega^2(2g_2+\lambda)(g_1+g_2+\lambda) = 
	\\ \nonumber
=(2g_1+\lambda)(2g_2+\lambda)((g_1+g_2+\lambda)^2+\Delta\omega^2)+4\Omega^2(g_1+g_2+\lambda)^2.
\end{gather}

If $\Delta\omega=0$, then
\begin{equation}
(g_1+g_2+\lambda)^2\left(((2g_1+\lambda)((2g_2+\lambda)+4\Omega^2\right)=0,
\end{equation}
\begin{equation}
\lambda_1=-(g_1+g_2),\ \ \ \ \ \lambda_2=-(g_1+g_2),
\end{equation}
\begin{equation}
\lambda_\pm=-(g_1+g_2)\pm\sqrt{(g_1-g_2)^2+4\Omega^2}.
\end{equation}

\section{Derivation of Eqs. (\ref{Jstt})-(\ref{Wip}). Stationary energy flow in local and PS approaches}
In this appendix, we derive Eqs. (\ref{Jstt})-(\ref{Wip}).
We use next designations: $\langle {\hat \sigma _1^\dag {{\hat \sigma }_1}} \rangle  = x$, $\langle {\hat \sigma _2^\dag {{\hat \sigma }_2}} \rangle  = y$, $\langle {\hat \sigma _1^\dag {{\hat \sigma }_2}} \rangle  = c + ip$.
So, system (\ref{SSeq_LOC}) for the occupanceis and energy flow in local approach can be rewritten as follows
\begin{equation}\label{SySSst}
\left\{ {\begin{array}{*{20}{c}}
{ - 2{g_1}x + 2\Omega p =  - {G_1}({\omega _1})}\\
{ - 2{g_2}y - 2\Omega p =  - {G_2}({\omega _2})}\\
{ - \left( {{g_1} + {g_2}} \right)c - \Delta \omega p = 0}\\
{ - \Omega x + \Omega y + \Delta \omega c - \left( {{g_1} + {g_2}} \right)p = 0}
\end{array}} \right.
\end{equation}
From this system of equations it is seen that
\begin{equation}
c =  - \frac{{\Delta \omega }}{{{g_1} + {g_2}}}p,\ \ \ \ 
x = \frac{{2\Omega p + {G_1}({\omega _1})}}{{2{g_1}}},\ \ \ \ 
y = \frac{{ - 2\Omega p + {G_2}({\omega _2})}}{{2{g_2}}}.
\end{equation}
Inserting these equations into the forth equation of the last system in Eq. (\ref{SySSst}), one can get the equation for $p$
\begin{equation}
\Omega \left( { - \frac{{2\Omega p + {G_1}({\omega _1})}}{{2{g_1}}} + \frac{{ - 2\Omega p + {G_2}({\omega _2})}}{{2{g_2}}}} \right) - \frac{{{{(\Delta \omega )}^2}}}{{g{}_1 + {g_2}}}p - \left( {{g_1} + {g_2}} \right)p = 0.
\end{equation}
and, consequently,
\begin{equation}\label{p}
p = - f\frac{\Omega }{e + \Omega ^2},\ \ \  f=\cfrac{{{G_1}({\omega _1}){G_2}( - {\omega _2}) - {G_1}( - {\omega _1}){G_2}({\omega _2})}}{{4\left( {{g_1} + {g_2}} \right)}}, \ \ \  e= g_1g_2 + \cfrac{{{{(\Delta \omega )}^2}}}{{{{\left( {g_1 + {g_2}} \right)}^2}}}g_1g_2
\end{equation}
Using Eqs. (\ref{SySSst})-(\ref{p}), one can calculate stationary energy flow:
\begin{gather}
{J_1^{st}} =  - 2{\omega _1}{g_1}x + {\omega _1}{G_1}({\omega _1}) - 2\Omega {g_1}c = 
	\\ \nonumber
= - \cfrac{{\left( {{\omega _1}{g_2} + {\omega _2}{g_1}} \right){\gamma _1}({\omega _1}){\gamma _2}({\omega _2})}}{{2{{\left( {{g_1} + {g_2}} \right)}^2}}}\cfrac{{{e^{{\omega _1}/{T_1}}} - {e^{{\omega _2}/{T_2}}}}}{{\left( {{e^{{\omega _1}/{T_1}}} - 1} \right)\left( {{e^{{\omega _2}/{T_2}}} - 1} \right)}}\cfrac{{{\Omega ^2}}}{{g{}_1g{}_2 + \frac{{{{(\Delta \omega )}^2}}}{{{{\left( {g{}_1 + {g_2}} \right)}^2}}}g{}_1g{}_2 + {\Omega ^2}}}
\end{gather}
Here we use Kubo-Martin-Shwinger condition $G_j(\pm\omega)=\gamma_j(\omega)(n_j(\omega)+1/2\mp1/2)$, $n_j(\omega)=1/(\mathrm{exp}(\hbar\omega/kT_j)-1)$,  $\omega>0$.
Maximum of $J_1^{st}$ with respect to $\Omega$ is achieved at
\begin{equation}
\Omega_{max}=\sqrt {g_1 g_2\left( {1 + \frac{{{{(\Delta \omega )}^2}}}{{{{\left( {g_1 + {g_2}} \right)}^2}}}} \right)}.
\end{equation}

\section{Explicit expressions for  Eqs.~(\ref{SSeq_PS})-(\ref{SEQ_PS})}
In this Appendix, we obtain explicit expressions for coefficients $A,B,C,F,T,Q,Y$ in Eqs.~(\ref{SSeq_PS})-(\ref{SEQ_PS}) and coefficients $A_G,B_G,C_G,F_G,T_G,Q_G,Y_G$ in Eqs.~(\ref{SSeq_GLB})-(\ref{SEQ_GLB}).
To do that, we introduce the following notation:
\begin{equation}
{\overrightarrow{G}_{GL1,2}} = \left( {\begin{array}{*{20}{c}}
{\frac{{{G_{1,2}}( - (\tilde \omega  + \tilde \Omega ))}}{2}}\\
{\frac{{{G_{1,2}}(\tilde \omega  + \tilde \Omega )}}{2}}\\
{\frac{{{G_{1,2}}( - (\tilde \omega  - \tilde \Omega ))}}{2}}\\
{\frac{{{G_{1,2}}(\tilde \omega  - \tilde \Omega )}}{2}}
\end{array}} \right),\ 
{\overrightarrow{G}_{PS1,2}} = \left( {\begin{array}{*{20}{c}}
{\frac{{{G_{1,2 - }}( - (\tilde \omega  - \tilde \Omega )) + {G_{1,2 + }}( - (\tilde \omega  + \tilde \Omega ))}}{2}}\\
{\frac{{{G_{1,2 - }}(\tilde \omega  + \tilde \Omega ) + {G_{1,2 + }}(\tilde \omega  - \tilde \Omega )}}{2}}\\
{\frac{{{G_{1,2 - }}(\tilde \omega  - \tilde \Omega ) + {G_{1,2 + }}(\tilde \omega  + \tilde \Omega )}}{2}}\\
{\frac{{{G_{1,2 - }}( - (\tilde \omega  + \tilde \Omega )) + {G_{1,2 + }}( - (\tilde \omega  - \tilde \Omega ))}}{2}}
\end{array}} \right)
\end{equation}
\begin{figure}
\includegraphics[width=0.5\linewidth]{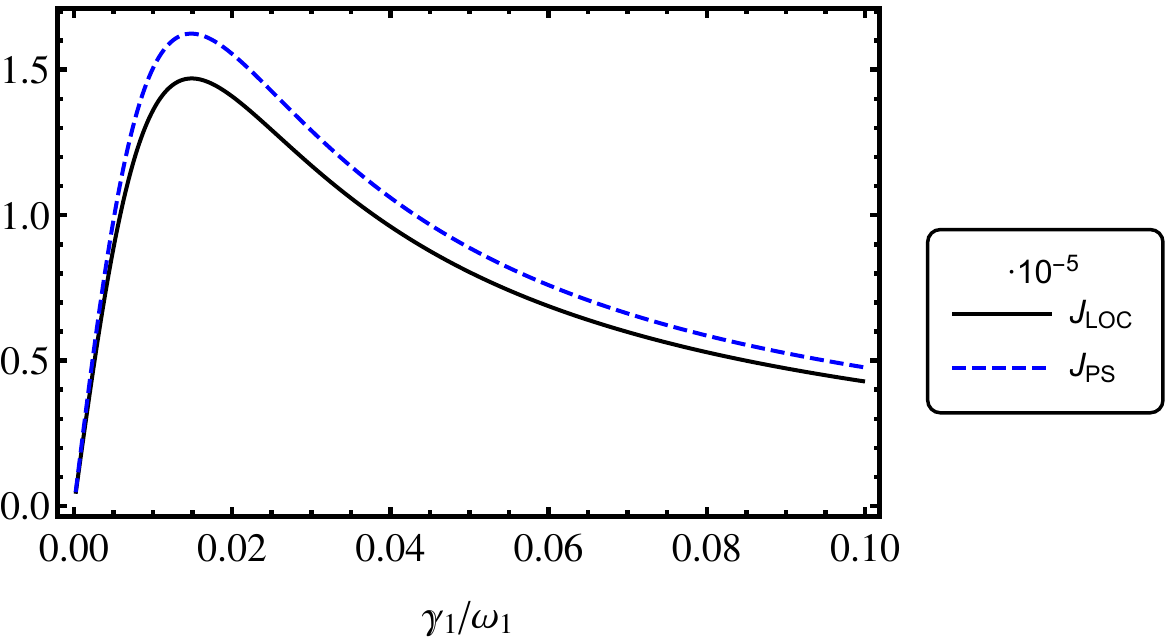}
\caption{Stationary energy flow $J^{st}$ to the hot reservoir in the local and the PS approaches as function of $\gamma_1$, $\omega_1=1.0$, $\gamma_k(\omega)=c_k\omega^3$,  $c_1=c_2/2$, $T_1=0.2$, $T_2=0.22$, $\Delta \omega=-0.009\omega_1$, $\Omega=0.01\Omega$.}
\label{LOC_PS_J}
\end{figure}

\begin{equation*}
{\overrightarrow{G}_{GL1,2\_FS}} = \left( {\begin{array}{*{20}{c}}
{\frac{{{G_{1,2 - }}( - (\tilde \omega  + \tilde \Omega )) - {G_{1,2 + }}( - (\tilde \omega  + \tilde \Omega ))}}{2}}\\
{\frac{{{G_{1,2 - }}(\tilde \omega  + \tilde \Omega ) - {G_{1,2 + }}(\tilde \omega  + \tilde \Omega )}}{2}}\\
{\frac{{{G_{1,2 - }}( - (\tilde \omega  - \tilde \Omega )) - {G_{1,2 + }}( - (\tilde \omega  - \tilde \Omega ))}}{2}}\\
{\frac{{{G_{1,2 - }}(\tilde \omega  - \tilde \Omega ) - {G_{1,2 + }}(\tilde \omega  - \tilde \Omega )}}{2}}
\end{array}} \right),\ 
{\overrightarrow{G}_{PS1,2\_FS}} = \left( {\begin{array}{*{20}{c}}
{\frac{{{G_{1,2 - }}( - (\tilde \omega  - \tilde \Omega )) - {G_{1,2 + }}( - (\tilde \omega  + \tilde \Omega ))}}{2}}\\
{\frac{{{G_{1,2 - }}(\tilde \omega  + \tilde \Omega ) - {G_{1,2 + }}(\tilde \omega  - \tilde \Omega )}}{2}}\\
{\frac{{{G_{1,2 - }}(\tilde \omega  - \tilde \Omega ) - {G_{1,2 + }}(\tilde \omega  + \tilde \Omega )}}{2}}\\
{\frac{{{G_{1,2 - }}( - (\tilde \omega  + \tilde \Omega )) - {G_{1,2 + }}( - (\tilde \omega  - \tilde \Omega ))}}{2}}
\end{array}} \right).
\end{equation*}

\begin{equation}
{\overrightarrow {\sigma _1^\dag {\sigma _1},\sigma _1^\dag {\sigma _1}} _{GL1}} = \left( {\begin{array}{*{20}{c}}
{ - 2{{\left( {\frac{{y + 2r}}{{4r}}} \right)}^2}}\\
{ - 2{{\left( {\frac{{y + 2r}}{{4r}}} \right)}^2}}\\
{ - 2{{\left( {\frac{{y - 2r}}{{4r}}} \right)}^2}}\\
{ - 2{{\left( {\frac{{y - 2r}}{{4r}}} \right)}^2}}
\end{array}} \right),\ \ \ 
{\overrightarrow {\sigma _1^\dag {\sigma _1},\sigma _1^\dag {\sigma _1}} _{PS1}} = \left( {\begin{array}{*{20}{c}}
{\frac{{{y^2} - 4{r^2}}}{{8{r^2}}}}\\
{\frac{{{y^2} - 4{r^2}}}{{8{r^2}}}}\\
{\frac{{{y^2} - 4{r^2}}}{{8{r^2}}}}\\
{\frac{{{y^2} - 4{r^2}}}{{8{r^2}}}}
\end{array}} \right),
\end{equation}

\begin{equation*}
{\overrightarrow {\sigma _1^\dag {\sigma _1},\sigma _1^\dag {\sigma _1}} _{GL2}} = \left( {\begin{array}{*{20}{c}}
{ - \frac{1}{{2{r^2}}}}\\
{ - \frac{1}{{2{r^2}}}}\\
{ - \frac{1}{{2{r^2}}}}\\
{ - \frac{1}{{2{r^2}}}}
\end{array}} \right),\ \ \ 
{\overrightarrow {\sigma _1^\dag {\sigma _1},\sigma _1^\dag {\sigma _1}} _{PS2}} = \left( {\begin{array}{*{20}{c}}
{\frac{1}{{2{r^2}}}}\\
{\frac{1}{{2{r^2}}}}\\
{\frac{1}{{2{r^2}}}}\\
{\frac{1}{{2{r^2}}}}
\end{array}} \right),
\end{equation*}

\begin{equation*}
{\overrightarrow {\sigma _1^\dag {\sigma _1},\sigma _1^\dag {\sigma _2}} _{GL1}} = \left( {\begin{array}{*{20}{c}}
{ - \frac{{y + 2r}}{{8{r^2}}}}\\
{ - \frac{{y + 2r}}{{8{r^2}}}}\\
{ - \frac{{y - 2r}}{{8{r^2}}}}\\
{ - \frac{{y - 2r}}{{8{r^2}}}}
\end{array}} \right),\ 
{\overrightarrow {\sigma _1^\dag {\sigma _1},\sigma _1^\dag {\sigma _2}} _{PS1}} = \left( {\begin{array}{*{20}{c}}
{\frac{{y - 2r}}{{8{r^2}}}}\\
{\frac{{y - 2r}}{{8{r^2}}}}\\
{\frac{{y + 2r}}{{8{r^2}}}}\\
{\frac{{y + 2r}}{{8{r^2}}}}
\end{array}} \right),\ 
{\overrightarrow {\sigma _1^\dag {\sigma _1},\sigma _1^\dag {\sigma _2}} _{GL2}} = \left( {\begin{array}{*{20}{c}}
{\frac{{y - 2r}}{{8{r^2}}}}\\
{\frac{{y - 2r}}{{8{r^2}}}}\\
{\frac{{y + 2r}}{{8{r^2}}}}\\
{\frac{{y + 2r}}{{8{r^2}}}}
\end{array}} \right),\ 
{\overrightarrow {\sigma _1^\dag {\sigma _1},\sigma _1^\dag {\sigma _2}} _{PS2}} = \left( {\begin{array}{*{20}{c}}
{\frac{{ - y + 2r}}{{8{r^2}}}}\\
{\frac{{ - y + 2r}}{{8{r^2}}}}\\
{ - \frac{{y + 2r}}{{8{r^2}}}}\\
{ - \frac{{y + 2r}}{{8{r^2}}}}
\end{array}} \right)
\end{equation*}

\begin{equation*}
{\overrightarrow {\sigma _1^\dag {\sigma _1},\sigma _1^\dag {\sigma _2}} _{GL1\_FS}} = \left( {\begin{array}{*{20}{c}}
{\frac{{y + 2r}}{{8{r^2}}}}\\
{ - \frac{{y + 2r}}{{8{r^2}}}}\\
{\frac{{y - 2r}}{{8{r^2}}}}\\
{ - \frac{{y - 2r}}{{8{r^2}}}}
\end{array}} \right),\ \ \ 
{\overrightarrow {\sigma _1^\dag {\sigma _1},\sigma _1^\dag {\sigma _2}} _{PS1\_FS}} = \left( {\begin{array}{*{20}{c}}
{ - \frac{{y - 2r}}{{8{r^2}}}}\\
{\frac{{y - 2r}}{{8{r^2}}}}\\
{\frac{{y + 2r}}{{8{r^2}}}}\\
{ - \frac{{y + 2r}}{{8{r^2}}}}
\end{array}} \right),
\end{equation*}

\begin{equation*}
{\overrightarrow {\sigma _1^\dag {\sigma _1},\sigma _1^\dag {\sigma _2}} _{GL2\_FS}} = \left( {\begin{array}{*{20}{c}}
{\frac{{ - y + 2r}}{{8{r^2}}}}\\
{ - \frac{{ - y + 2r}}{{8{r^2}}}}\\
{ - \frac{{y + 2r}}{{8{r^2}}}}\\
{\frac{{y + 2r}}{{8{r^2}}}}
\end{array}} \right),\ \ \ 
{\overrightarrow {\sigma _1^\dag {\sigma _1},\sigma _1^\dag {\sigma _2}} _{PS2\_FS}} = \left( {\begin{array}{*{20}{c}}
{\frac{{y - 2r}}{{8{r^2}}}}\\
{ - \frac{{y - 2r}}{{8{r^2}}}}\\
{ - \frac{{y + 2r}}{{8{r^2}}}}\\
{\frac{{y + 2r}}{{8{r^2}}}}
\end{array}} \right).
\end{equation*}

\begin{equation}
{\overrightarrow {\sigma _2^\dag {\sigma _2},\sigma _2^\dag {\sigma _2}} _{GL1}} ={\overrightarrow{\sigma _1^\dag {\sigma _1},\sigma _1^\dag {\sigma _1}} _{GL2}},\ \ \ 
{\overrightarrow {\sigma _2^\dag {\sigma _2},\sigma _2^\dag {\sigma _2}} _{PS1}} = {\overrightarrow{\sigma _1^\dag {\sigma _1},\sigma _1^\dag {\sigma _1}} _{PS2}},
\end{equation}

\begin{equation*}
{\overrightarrow {\sigma _2^\dag {\sigma _2},\sigma _2^\dag {\sigma _2}} _{GL2}} = \left( {\begin{array}{*{20}{c}}
{ - 2{{\left( {\frac{{ - y + 2r}}{{4r}}} \right)}^2}}\\
{ - 2{{\left( {\frac{{ - y + 2r}}{{4r}}} \right)}^2}}\\
{ - 2{{\left( {\frac{{y + 2r}}{{4r}}} \right)}^2}}\\
{ - 2{{\left( {\frac{{y + 2r}}{{4r}}} \right)}^2}}
\end{array}} \right),\ \ \ 
{\overrightarrow {\sigma _2^\dag {\sigma _2},\sigma _2^\dag {\sigma _2}} _{PS2}} = \left( {\begin{array}{*{20}{c}}
{\frac{{{y^2} - 4{r^2}}}{{8{r^2}}}}\\
{\frac{{{y^2} - 4{r^2}}}{{8{r^2}}}}\\
{\frac{{{y^2} - 4{r^2}}}{{8{r^2}}}}\\
{\frac{{{y^2} - 4{r^2}}}{{8{r^2}}}}
\end{array}} \right)
\end{equation*}

\begin{equation*}
{\overrightarrow {\sigma _2^\dag {\sigma _2},\sigma _1^\dag {\sigma _2}} _{GL1}} ={\overrightarrow{\sigma _1^\dag {\sigma _1},\sigma _1^\dag {\sigma _2}} _{GL1}},\ 
{\overrightarrow {\sigma _2^\dag {\sigma _2},\sigma _1^\dag {\sigma _2}} _{PS1}} ={\overrightarrow{\sigma _1^\dag {\sigma _1},\sigma _1^\dag {\sigma _2}} _{PS1}},
\end{equation*}
\begin{equation*}
{\overrightarrow {\sigma _2^\dag {\sigma _2},\sigma _1^\dag {\sigma _2}} _{GL2}} = {\overrightarrow{\sigma _1^\dag {\sigma _1},\sigma _1^\dag {\sigma _2}} _{GL2}},\ 
{\overrightarrow {\sigma _2^\dag {\sigma _2},\sigma _1^\dag {\sigma _2}} _{PS2}} = {\overrightarrow{\sigma _1^\dag {\sigma _1},\sigma _1^\dag {\sigma _2}} _{PS2}},
\end{equation*}

\begin{equation*}
{\overrightarrow {\sigma _2^\dag {\sigma _2},\sigma _1^\dag {\sigma _2}} _{GL1\_FS}} =  -{\overrightarrow{\sigma _1^\dag {\sigma _1},\sigma _1^\dag {\sigma _2}} _{GL1\_FS}},\ 
{\overrightarrow {\sigma _2^\dag {\sigma _2},\sigma _1^\dag {\sigma _2}} _{PS1\_FS}} =  - {\overrightarrow{\sigma _1^\dag {\sigma _1},\sigma _1^\dag {\sigma _2}} _{PS1\_FS}},
\end{equation*}

\begin{equation*}
{\overrightarrow {\sigma _2^\dag {\sigma _2},\sigma _1^\dag {\sigma _2}} _{GL2\_FS}} =  - {\overrightarrow{\sigma _1^\dag {\sigma _1},\sigma _1^\dag {\sigma _2}} _{GL2\_FS}},\ 
{\overrightarrow {\sigma _2^\dag {\sigma _2},\sigma _1^\dag {\sigma _2}} _{PS2\_FS}} =  - {\overrightarrow{\sigma _1^\dag {\sigma _1},\sigma _1^\dag {\sigma _2}} _{PS2\_FS}}.
\end{equation*}

\begin{equation}
{\overrightarrow {\sigma _1^\dag {\sigma _2},\sigma _1^\dag {\sigma _2}} _{GL1\_FS}} = \left( {\begin{array}{*{20}{c}}
{ - {{\left( {\frac{{y + 2r}}{{4r}}} \right)}^2} + \frac{1}{{4{r^2}}}}\\
{{{\left( {\frac{{y + 2r}}{{4r}}} \right)}^2} - \frac{1}{{4{r^2}}}}\\
{ - {{\left( {\frac{{y - 2r}}{{4r}}} \right)}^2} + \frac{1}{{4{r^2}}}}\\
{{{\left( {\frac{{y - 2r}}{{4r}}} \right)}^2} - \frac{1}{{4{r^2}}}}
\end{array}} \right),\ 
{\overrightarrow {\sigma _1^\dag {\sigma _2},\sigma _1^\dag {\sigma _2}} _{PS1\_FS}} = \left( {\begin{array}{*{20}{c}}
{\frac{{{y^2} - 4{r^2}}}{{16{r^2}}} - \frac{1}{{4{r^2}}}}\\
{ - \frac{{{y^2} - 4{r^2}}}{{16{r^2}}} + \frac{1}{{4{r^2}}}}\\
{ - \frac{{{y^2} - 4{r^2}}}{{16{r^2}}} + \frac{1}{{4{r^2}}}}\\
{\frac{{{y^2} - 4{r^2}}}{{16{r^2}}} - \frac{1}{{4{r^2}}}}
\end{array}} \right)
\end{equation}

\begin{equation*}
{\overrightarrow {\sigma _1^\dag {\sigma _2},\sigma _1^\dag {\sigma _2}} _{GL2\_FS}} = \left( {\begin{array}{*{20}{c}}
{{{\left( {\frac{{ - y + 2r}}{{4r}}} \right)}^2} - \frac{1}{{4{r^2}}}}\\
{ - {{\left( {\frac{{ - y + 2r}}{{4r}}} \right)}^2} + \frac{1}{{4{r^2}}}}\\
{{{\left( {\frac{{y + 2r}}{{4r}}} \right)}^2} - \frac{1}{{4{r^2}}}}\\
{ - {{\left( {\frac{{y + 2r}}{{4r}}} \right)}^2} + \frac{1}{{4{r^2}}}}
\end{array}} \right),\ 
{\overrightarrow {\sigma _1^\dag {\sigma _2},\sigma _1^\dag {\sigma _2}} _{PS2\_FS}} = \left( {\begin{array}{*{20}{c}}
{ - \frac{{{y^2} - 4{r^2}}}{{16{r^2}}} + \frac{1}{{4{r^2}}}}\\
{\frac{{{y^2} - 4{r^2}}}{{16{r^2}}} - \frac{1}{{4{r^2}}}}\\
{\frac{{{y^2} - 4{r^2}}}{{16{r^2}}} - \frac{1}{{4{r^2}}}}\\
{ - \frac{{{y^2} - 4{r^2}}}{{16{r^2}}} + \frac{1}{{4{r^2}}}}
\end{array}} \right)
\end{equation*}
where $y = \left( {{\omega _1} - {\omega _2}} \right)/\Omega $, $r = \sqrt {{y^2}/4 + 1} $.
Using these notations, the coefficients in Eqs.~(\ref{SSeq_PS})-(\ref{SEQ_PS}) and Eqs.~(\ref{SSeq_GLB})-(\ref{SEQ_GLB}) can be written in the following form:
\begin{equation}
A = A_1 + A_2,\ \ \ \ 
A_j = A_{GLj} + A_{PSj},\ \ \ \ 
{A_{GLj}} = \left( {{{\overrightarrow{G}}_{GLj}},{{\overrightarrow {\sigma _1^\dag {\sigma _1},\sigma _1^\dag {\sigma _1}} }_{GLj}}} \right),\ \ \ \ 
{A_{PSj}} = \left( {{{\overrightarrow{G}}_{PSj}},{{\overrightarrow {\sigma _1^\dag {\sigma _1},\sigma _1^\dag {\sigma _1}} }_{PSj}}} \right),
\end{equation}
\begin{equation}
F = F_1 + F_2,\ \ \ \ 
F_j = F_{GLj} + F_{PSj},\ \ \ \ 
{F_{GLj}} = \left( {{{\overrightarrow{G}}_{GLj}},{{\overrightarrow {\sigma _2^\dag {\sigma _2},\sigma _2^\dag {\sigma _2}} }_{GLj}}} \right),\ \ \ \ 
{F_{PSj}} = \left( {{{\overrightarrow{G}}_{PSj}},{{\overrightarrow {\sigma _2^\dag {\sigma _2},\sigma _2^\dag {\sigma _2}} }_{PSj}}} \right),
\end{equation}
\begin{equation*}
B = {B_1} + {B_2},\ \ \ \ 
{B_j} = {B_{GLj}} + {B_{PSj}},\ \ \ \ 
{B_{GLj}} = \left( {{{\overrightarrow{G}}_{GLj}},{{\overrightarrow {\sigma _1^\dag {\sigma _1},\sigma _1^\dag {\sigma _2}} }_{GLj}}} \right),\ \ \ \ 
{B_{PSj}} = \left( {{{\overrightarrow{G}}_{PSj}},{{\overrightarrow {\sigma _1^\dag {\sigma _1},\sigma _1^\dag {\sigma _2}} }_{PSj}}} \right),
\end{equation*}
\begin{equation*}
C = {C_1} + {C_2},\ \ \ \ 
{C_j} = {C_{GLj}} + {C_{PSj}},\ \ \ \ 
{C_{GLj}} = \left( {{{\overrightarrow{G}}_{GLj\_FS}},{{\overrightarrow {\sigma _1^\dag {\sigma _1},\sigma _1^\dag {\sigma _2}} }_{GLj\_FS}}} \right)\ \ \ \ 
{C_{PSj}} = \left( {{{\overrightarrow{G}}_{PSj\_FS}},{{\overrightarrow {\sigma _1^\dag {\sigma _1},\sigma _1^\dag {\sigma _2}} }_{PSj\_FS}}} \right)
\end{equation*}
\begin{equation*}
D=D_1+D_2,\ \ \ \ 
{D_j} = {D_{GLj}} + {D_{PSj}},\ \ \ \ 
D_{GLj}=({{\overrightarrow{G}}_{GLj\_FS}},{\overrightarrow {\sigma _1^\dag {\sigma _2},\sigma _1^\dag {\sigma _2}} _{GLj\_FS}}),\ \ \ \ 
({{\overrightarrow{G}}_{PSj\_FS}},{\overrightarrow {\sigma _1^\dag {\sigma _2},\sigma _1^\dag {\sigma _2}} _{PSj\_FS}})
\end{equation*}


To write down the explicit expression for free terms in Eqs.~(\ref{SSeq_PS})-(\ref{SEQ_PS}) and Eqs.~(\ref{SSeq_GLB})-(\ref{SEQ_GLB}), it is convenient to introduce the following notations:
\begin{equation}
{\overrightarrow {\sigma _1^\dag {\sigma _1}} _{GL1\_S}} = \left( {\begin{array}{*{20}{c}}
0\\
{2{{\left( {\frac{{y + 2r}}{{4r}}} \right)}^2}}\\
0\\
{2{{\left( {\frac{{y - 2r}}{{4r}}} \right)}^2}}
\end{array}} \right),\ 
{\overrightarrow {\sigma _1^\dag {\sigma _1}} _{PS1\_S}} = \left( {\begin{array}{*{20}{c}}
0\\
{ - \frac{{{y^2} - 4{r^2}}}{{8{r^2}}}}\\
{ - \frac{{{y^2} - 4{r^2}}}{{8{r^2}}}}\\
0
\end{array}} \right),\ 
{\overrightarrow {\sigma _1^\dag {\sigma _1}} _{GL2\_S}} = \left( {\begin{array}{*{20}{c}}
0\\
{\frac{1}{{2{r^2}}}}\\
0\\
{\frac{1}{{2{r^2}}}}
\end{array}} \right),\ 
{\overrightarrow {\sigma _1^\dag {\sigma _1}} _{PS2\_S}} = \left( {\begin{array}{*{20}{c}}
0\\
{ - \frac{1}{{2{r^2}}}}\\
{ - \frac{1}{{2{r^2}}}}\\
0
\end{array}} \right),
\end{equation}

\begin{equation*}
{\overrightarrow {\sigma _2^\dag {\sigma _2}} _{GL1\_S}} = \left( {\begin{array}{*{20}{c}}
0\\
{\frac{1}{{2{r^2}}}}\\
0\\
{\frac{1}{{2{r^2}}}}
\end{array}} \right),\ 
{\overrightarrow {\sigma _2^\dag {\sigma _2}} _{PS1\_S}} = \left( {\begin{array}{*{20}{c}}
0\\
{ - \frac{1}{{2{r^2}}}}\\
{ - \frac{1}{{2{r^2}}}}\\
0
\end{array}} \right),\ 
{\overrightarrow {\sigma _2^\dag {\sigma _2}} _{GL2\_S}} = \left( {\begin{array}{*{20}{c}}
0\\
{2{{\left( {\frac{{ - y + 2r}}{{4r}}} \right)}^2}}\\
0\\
{2{{\left( {\frac{{y + 2r}}{{4r}}} \right)}^2}}
\end{array}} \right),\ 
{\overrightarrow {\sigma _2^\dag {\sigma _2}} _{PS2\_S}} = \left( {\begin{array}{*{20}{c}}
0\\
{ - \frac{{{y^2} - 4{r^2}}}{{8{r^2}}}}\\
{ - \frac{{{y^2} - 4{r^2}}}{{8{r^2}}}}\\
0
\end{array}} \right),
\end{equation*}

\begin{equation*}
{\overrightarrow {\sigma _1^\dag {\sigma _2}} _{GL1\_S}} = \left( {\begin{array}{*{20}{c}}
0\\
{\frac{{y + 2r}}{{4{r^2}}}}\\
0\\
{\frac{{y - 2r}}{{4{r^2}}}}
\end{array}} \right),\ 
{\overrightarrow {\sigma _1^\dag {\sigma _2}} _{PS1\_S}} =  - \left( {\begin{array}{*{20}{c}}
0\\
{\frac{{y + 2r}}{{4{r^2}}}}\\
{\frac{{y - 2r}}{{4{r^2}}}}\\
0
\end{array}} \right),\ 
{\overrightarrow {\sigma _1^\dag {\sigma _2}} _{GL2\_S}} = \left( {\begin{array}{*{20}{c}}
0\\
{\frac{{ - y + 2r}}{{4{r^2}}}}\\
0\\
{ - \frac{{y + 2r}}{{4{r^2}}}}
\end{array}} \right),\ 
{\overrightarrow {\sigma _1^\dag {\sigma _2}} _{PS2\_S}} = \left( {\begin{array}{*{20}{c}}
0\\
{\frac{{y + 2r}}{{4{r^2}}}}\\
{\frac{{y - 2r}}{{4{r^2}}}}\\
0
\end{array}} \right).
\end{equation*}
Using these notations, one can write
\begin{equation}
{S_{11}} = ({\overrightarrow{G}_{GL1}},{\overrightarrow {\sigma _1^\dag {\sigma _1}} _{GL1\_S}}) + ({ \overrightarrow{G}_{PS1}},{\overrightarrow{\sigma _1^\dag {\sigma _1}} _{PS1\_S}}) + ({\overrightarrow{G}_{GL2}},{\overrightarrow{\sigma _1^\dag {\sigma _1}} _{GL2\_S}}) + ({\overrightarrow{G}_{PS2}},{\overrightarrow{\sigma _1^\dag {\sigma _1}} _{PS2\_S}}),\ 
\end{equation}

\begin{equation*}
{S_{22}} = ({\overrightarrow{G}_{GL1}},{\overrightarrow {\sigma _2^\dag {\sigma _2}} _{GL1\_S}}) + ({\overrightarrow{G}_{PS1}},{\overrightarrow{\sigma _2^\dag {\sigma _2}} _{PS1\_S}}) + ({\overrightarrow{G}_{GL2}},{\overrightarrow{\sigma _2^\dag {\sigma _2}} _{GL2\_S}}) + ({\overrightarrow{G}_{PS2}},{\overrightarrow{\sigma _2^\dag {\sigma _2}} _{PS2\_S}})
\end{equation*}

\begin{equation*}
{S_{12}} = ({\overrightarrow{G}_{GL1}},{\overrightarrow {\sigma _1^\dag {\sigma _2}} _{GL1\_S}}) + ({\overrightarrow{G}_{PS1}},{\overrightarrow{\sigma _1^\dag {\sigma _2}} _{PS1\_S}}) + ({\overrightarrow{G}_{GL2}},{\overrightarrow{\sigma _1^\dag {\sigma _2}} _{GL2\_S}}) + ({\overrightarrow{G}_{PS2}},{\overrightarrow{\sigma _1^\dag {\sigma _2}} _{PS2\_S}}).
\end{equation*}
To obtain the system (\ref{SEQ_GLB}) for occupancies in the global approach, one should consider $A_{PSj}=B_{PSj}=C_{PSj}=D_{PSj}=0$.
Frequency shifts are neglected in the main text, so $C_{GLj}=D_{GLj}=0$ are considered in both global Eq.~(\ref{SEQ_GLB}) and PS Eq.~(\ref{SEQ_PS}) approaches.

\section{Stationary energy flow in the global approach}
In this Appendix, we derive explicit expressions for the energy flow in the global approach for the case $\Delta \omega=0$.
To do this, it is convenient to rewrite the Hamiltonian of two coupled TLSs with same frequency in the form
\begin{equation}
{\hat H_S} = \omega\hat \sigma _1^\dag {\hat \sigma _1} + {\omega}\hat \sigma _2^\dag {\hat \sigma _2} + \Omega \left( {\hat \sigma _1^\dag {{\hat \sigma }_2} + \hat \sigma _2^\dag {{\hat \sigma }_1}} \right)=(\omega+\Omega)\hat A^\dag \hat A+(\omega-\Omega)\hat B^\dag \hat  B.
\end{equation}
Here $\hat A=\sqrt{2}((\hat \sigma_1+\hat \sigma_2)/2-\hat \sigma^\dag_1\hat \sigma_1\hat \sigma_2)$, $B=\sqrt{2}((\hat \sigma_1-\hat \sigma_2)/2+\hat \sigma^\dag_1\hat \sigma_1\hat \sigma_2)$.
One can see that
\begin{equation}
\overrightarrow{A^\dag A}=
\left( {\begin{array}{*{20}{c}}
{{\hat A^\dag }\hat A}\\
{{\hat B^\dag }\hat B}\\
{{\hat A^\dag }\hat B}\\
{{\hat B^\dag }\hat A}
\end{array}} \right) = \frac{1}{2}\left( {\begin{array}{*{20}{c}}
1&1&1&1\\
1&1&{ - 1}&{ - 1}\\
1&{ - 1}&{ - 1}&1\\
1&{ - 1}&1&{ - 1}
\end{array}} \right) \times \left( {\begin{array}{*{20}{c}}
{\hat \sigma_1^\dag {\hat \sigma_1}}\\
{\hat \sigma_2^\dag {\hat \sigma_2}}\\
{\hat \sigma_1^\dag {\hat \sigma_2}}\\
{\hat \sigma_2^\dag {\hat \sigma_1}}
\end{array}} \right) = Q \times \left( {\begin{array}{*{20}{c}}
{\hat \sigma_1^\dag {\hat \sigma_1}}\\
{\hat \sigma_2^\dag {\hat \sigma_2}}\\
{\hat \sigma_1^\dag {\hat \sigma_2}}\\
{\hat \sigma_2^\dag {\hat \sigma_1}}
\end{array}} \right),\ \ \ \ \
Q^{-1}=Q.
\end{equation}
We can rewrite the system (\ref{SSeq_GLB}) in the terms of occupancies of eigenstates of the Hamiltonian
\begin{equation}
\frac{{d\overrightarrow{\left\langle {{{ \sigma }^\dag } \sigma } \right\rangle} }}{{dt}} = {M_{\rm{G} }}\overline{\left\langle {{{ \sigma }^\dag } \sigma } \right\rangle}  + {\overline{G}_{\rm{G} }}
\end{equation}
\begin{equation*}
Q\frac{{d\overrightarrow{\left\langle {{{ \sigma }^\dag } \sigma } \right\rangle} }}{{dt}} =Q {M_{\rm{G} }}Q^{-1}Q\overline{\left\langle {{{ \sigma }^\dag } \sigma } \right\rangle}  +Q {\overline{G}_{\rm{G} }}
\end{equation*}
\begin{equation*}
\frac{{d\overrightarrow{\left\langle {{{ A }^\dag } A } \right\rangle} }}{{dt}} = {M_{\rm{G,A} }}\overline{\left\langle {{{ A }^\dag } A } \right\rangle}  + {\overline{G}_{\rm{G,A} }}
\end{equation*}
Thus one can obtain (as it follows from the Appendix C)
\begin{equation}
M_G=\left( {\begin{array}{*{20}{c}}
q&0&{ - i\Omega  + v}&{i\Omega  + v }\\
0&q&{i\Omega  + v}&{ - i\Omega  + v }\\
{ - i\Omega  + v }&{i\Omega  + v }&{q}&0\\
{i\Omega  +v}&{ - i\Omega  +v}&0&{ q}
\end{array}} \right)\Rightarrow
M_{G,A}=\left( {\begin{array}{*{20}{c}}
q+2v&0&0&0\\
0&q-2v&0&0\\
0&0&{q+2i\Omega}&0\\
0&0&0&{ q -2i\Omega}
\end{array}} \right)
\end{equation}
Here $q=-(g_s+g_a)/2$, $v=(-g_s+g_a)/4$, $g_s=(g_1(\omega+\Omega)+g_2(\omega+\Omega))/2$, $g_a=(g_1(\omega-\Omega)+g_2(\omega-\Omega))/2$,   ${g_j}(\omega) = ({{G_j}( - {\omega}) + {G_j}({\omega})})/2=\gamma_j(\omega)(n_j(\omega)+1/2)$.

\begin{equation}\label{SysGlob}
M_{G,A}=\left( {\begin{array}{*{20}{c}}
-g_s&0&0&0\\
0&-g_a&0&0\\
0&0&{-g_s/2-g_a/2+2i\Omega}&0\\
0&0&0&{ -g_s/2-g_a/2 -2i\Omega}
\end{array}} \right),\ \ \ \ \ 
\overrightarrow{G}_{\rm{G,A}}=\frac{1}{2}
 \left( {\begin{array}{*{20}{c}}
{G_1(\omega+\Omega)+G_2(\omega+\Omega)}\\
{G_1(\omega-\Omega)+G_2(\omega-\Omega)}\\
{0}\\
{0}
\end{array}} \right).
\end{equation}

The stationary solution of Eq.~(\ref{SysGlob}) is 
\begin{equation}
\overrightarrow{\langle A^\dag A\rangle}=
 \left( {\begin{array}{*{20}{c}}
{\cfrac{n_{1s}\gamma_{1s}+n_{2s}\gamma_{2s}}{(2n_{1s}+1)\gamma_{1s}+(2n_{2s}+1)\gamma_{2s}}}\\
{\cfrac{n_{1a}\gamma_{1a}+n_{2a}\gamma_{2a}}{(2n_{1a}+1)\gamma_{1a}+(2n_{2a}+1)\gamma_{2a}}}\\
{0}\\
{0}
\end{array}} \right),
\ \ \ \ \ 
{\begin{array}{*{20}{c}}
n_{1,2s}=n_{1,2}(\omega+\Omega)\\
n_{1,2a}=n_{1,2}(\omega-\Omega)\\
\gamma_{1,2s}=\gamma_{1,2}(\omega+\Omega)\\
\gamma_{1,2a}=\gamma_{1,2}(\omega-\Omega)
\end{array}}.
\end{equation}

The stationary energy flow is
\begin{align}
J_1^{st}=\frac{(\omega+\Omega)}{2}\cfrac{\gamma_{1s}\gamma_{2s}(n_{1s}-n_{2s})}{(2n_{1s}+1)\gamma_{1s}+(2n_{2s}+1)\gamma_{2s}}+\frac{(\omega-\Omega)}{2}\cfrac{\gamma_{1a}\gamma_{2a}(n_{1a}-n_{2a})}{(2n_{1a}+1)\gamma_{1a}+(2n_{2a}+1)\gamma_{2a}}
\end{align}

In the case $\gamma_{j}=c_j \omega^n$, we have
\begin{equation}
J_1^{st}=\frac{(\omega+\Omega)^{n+1}}{2}\cfrac{c_1 c_2(n_{1s}-n_{2s})}{(2n_{1s}+1)c_1+(2n_{2s}+1)c_2}+\frac{(\omega-\Omega)^{n+1}}{2}\cfrac{c_1 c_2(n_{1a}-n_{2a})}{(2n_{1a}+1)c_1+(2n_{2a}+1)c_2}
\end{equation}
Without loss of the generality, we consider $T_1<T_2$.
When $\Omega<\omega$, thus, the stationary energy flow negative $J_1^{st}<0$.
If $c_1~c_2$, $J_1^{st}$ linearly grows along with $c_1$ and $\gamma_1$.

\end{widetext}

\end{document}